\documentstyle[multicol,eqsecnum,aps,pre,epsf]{revtex}
\begin{document}
\title{ Diffusion limited aggregation as a Markovian process: \\
	site-sticking conditions}
\author{Boaz Kol and Amnon Aharony}
\address{Raymond and Beverly Sackler Faculty of Exact Sciences, 
	School of Physics and Astronomy,\\ Tel Aviv University, 
	69978 Ramat Aviv, Israel}
\date{\today}
\maketitle
\begin{abstract}
Cylindrical lattice 
diffusion limited aggregation (DLA), with a narrow width $N$, 
is solved for site-sticking conditions
using a Markovian matrix method
(which was previously developed for the bond-sticking case).
This matrix contains the probabilities that the front moves from one
configuration to another at each growth step, calculated
exactly by solving the Laplace equation and using the proper normalization.
The method is applied for a series of approximations, which include
only a finite number of rows near the front.
The fractal
dimensionality of the aggregate is extrapolated to a value near
$1.68$. 
\end{abstract}
\widetext
\begin{multicols}{2}
\section{Introduction}
Diffusion limited aggregation (DLA) \cite{Witten83} has been the subject of 
extensive study since it was first introduced. This model exhibits a growth
process that produces highly ramified self similar patterns, which are 
believed to be fractals \cite{Mandelbrot82}. It seems that DLA captures the 
essential mechanism in many natural growth processes, such as viscous 
fingering \cite{Feder}, dielectric breakdown \cite{DBM84a,DBM84b}, etc.
In spite of
the apparent simplicity of the model, an analytic solution is still 
unavailable. Particularly, the exact value of the fractal dimension
is not known. Some of the analytic approaches employed so far include the
fixed scale transformation (FST) \cite{Erzan95}, real space renormalization 
group (RSRG) \cite{Gould86,Nagatani88,Wang89} and conformal mapping 
\cite{Hastings98,Hastings97,Davidovitch99,Davidovitch2000}.

In DLA there is a seed cluster of particles fixed somewhere. 
A particle is released at a distance from the cluster, and performs a random 
walk until it attempts to penetrate the fixed cluster, in which case it 
sticks. Then the next particle is released and so on. There are two common 
types of sticking conditions. The sticking condition described above
is called ``bond-DLA'', because it occurs when the random walker attempts to
cross a perimeter bond between an unoccupied site and the aggregate.
In an earlier paper \cite{Kol2000a}, we solved the bond-DLA problem using
a Markovian process. Here we apply similar methods to the ``site-DLA'' case,
where sticking occurs as soon as the random walker 
arrives at a site that is a nearest neighbor to the aggregate. 
Since it is believed that
the large scale
structure of DLA is not sensitive to the type of sticking conditions used 
\cite{Cafiero93,Erzan95}, one expects both problems to yield the same
asymptotic fractal structures.

DLA can be grown in various geometries.
In this paper we deal with the cylindrical geometry in two dimensions (2D),
where the particles are emitted from a 
distant horizontal line at the top, while the
seed cluster is a parallel line at the bottom, with periodic boundary 
conditions on the sides. We only consider relatively narrow 
cylinders, with widths ranging from $N=2$ to $N=10$.
Even though the analysis in this paper is solely 
2D, the same techniques can be applied in higher dimensions.

An exact solution of bond-DLA with $N=2$ was published in 1998 \cite{Kol98}.
A generalization of the same approach was used in order to solve slightly 
wider cases with $N$ between $3$ and $7$ \cite{Kol2000a}. 
The solution presented in the latter case is not exact, 
but still, it presents a well controlled series of approximations in the sense
that any desired numerical accuracy could be obtained, provided that
a sufficiently high-order of approximation is used. The difficulty with
performing a high-order calculation is that its complexity grows
exponentially. 

The main idea in these references and in this paper is to follow the dynamics
of the growing front. The shape of the interface determines the
unique solution to the Laplace equation that determines the growth 
probabilities. The structure of the aggregate behind the interface is 
irrelevant and so is the history that led to the 
current interface. Each growth process changes the interface. We can therefore
describe DLA as a Markovian flow in the space of interface configurations.
The Markov states are the possible shapes of the interface,
which are indexed by an integer, usually denoted by $i$ or $j$. 
$P_i(t+1)$, the 
probability that the interface is in state $i$ at time $t+1$, 
depends only on the state of the interface at time $t$. 
The conditional transition 
probabilities from state $j$ to state $i$ make up the evolution matrix 
$E_{i,j}$, which is time independent. Thus, 
the dynamics of the Markov chain is described by the Master equations, 
\begin{equation}
P_i(t+1)=\sum_jE_{i,j}P_j(t),~~i=1,2,\dots.
\label{Markov dynamics}
\end{equation}
Each matrix element $E_{i,j}$ corresponds to a particular growth process,
and the sum of $j$ runs over all the interface configurations (whose number
may be infinite).
In order to fully describe the dynamics, it is necessary to calculate
the probabilities of all the possible growth processes, for each
of the possible initial configurations of the interface. We calculate
the growth probabilities by solving the discrete Laplace equation
on a lattice for a function $\Phi$, which corresponds to the average density
of random walkers, 
\begin{equation}
\nabla^2\Phi=0.
\label{Laplace}
\end{equation}
In the dielectric breakdown model (DBM) \cite{DBM84a,DBM84b} 
$\Phi$ has an electrostatic meaning, so
it is also commonly referred to as the ``potential''. 

Usually, the equation set (\ref{Markov dynamics}) is infinite because the 
number of possible shapes the interface may assume is unlimited. This may
pose a problem for two reasons. For one, it is difficult to include all of the
possibilities systematically. 
The case of bond-DLA with $N=2$ is a counter-example, where the complete set
of possible configurations can be easily characterized using a single 
parameter. This is because the
interface has the shape of a step whose height $j$ can be any nonnegative 
integer \cite{Kol98}. 
For $N>2$, however, it
is very difficult to parametrise the shape of the interface, even with the
use of more than one parameter, because complex overhangs may occur 
\cite{Kol2000a}. 
The second problem is that even if it was possible to account for a complete 
infinite set of configurations, it would still be awkward to analyze
the Markov process, e.g., finding its fixed point. 
Instead of accounting for all the configurations,
we make an approximation by employing some consistent truncation scheme
on the list of configurations. 
In the $O$'th-order approximation we include
only the top $O$ rows of a configuration and truncate the rest;
The list of configurations is sorted according to the maximal height
difference, $\Delta m$, between the lowest and highest particles on the 
interface \cite{Kol2000a}. 
In the $O$'th-order approximation, only a finite set of 
configurations with $\Delta m\leq O$ are taken into account. The configurations
with $\Delta m>O$ are truncated so that only their top $O$ rows are taken into
account (below the $O$'th row all the sites are considered to be occupied).
This truncation does not have a noticeable effect on the upward growth
probability (the growth probability at the tip), because of the exponential
decay of the potential inside deep fjords. Because of this exponential decay,
the approximation converges very fast as a function of $O$. Unfortunately,
the number of configurations diverges exponentially with $O$, so that the
calculation can be carried out only for relatively low-order (depending
on the width $N$ and on the strength of the computer).

In the case of site-DLA, the situation is a bit simpler than in bond-DLA,
because it is generally
harder for the random walker to penetrate deep into a fjord. A particle will 
only be able to enter fjords that are three sites wide or more, unlike the
case of bond-DLA, where a particle can go into a single column fjord. This 
makes the solution of site-DLA with $N=2$ and $N=3$ much simpler,
because they both have only a finite number of interface configurations. 
The narrowest cylinder that can have an arbitrarily deep fjord 
(a configuration with an arbitrarily large $\Delta m$)
arises for $N=4$, and thus
there is an infinite number of configurations. However, there can be no 
fluctuations in the width of the fjord, so in this sense this case resembles
the $N=2$ case in bond-DLA. For $N>4$ the approximation method must be
used, but generally, for the same $N$ and $O$ the number of configurations in
site-DLA is much smaller than in bond-DLA, so it is possible to perform 
higher-order calculations for wider cylinders.

Once an order of approximation $O$ is chosen, there is only a finite
number of configurations, $N_c(N,O)$, which depends both on $N$ and on $O$.
The Markov process is then closed and irreducible. Closed
means that $\sum_{i=1}^{N_c}E_{i,j}=1$ for $j=1,\dots,N_c$, and irreducible 
means that there is a finite probability to go from
any initial state $j$ to any final state $i$ during a finite number of time 
steps. A basic theorem in Markov theory states that a closed and irreducible
process necessarily has a single fixed point \cite{Norris97}. 
This fixed point represents
the steady-state probabilities for the various interface configurations in
the asymptotic time limit. The  theorem is also true for an
infinite number of states, so one can conclude that the unapproximated 
process also converges to a steady state.
As mentioned, this steady state is characterized for example by a  
time independent average density $\rho$.

The fixed point equations are,
\begin{equation}
P^*_i=\sum_jE_{i,j}P^*_j,~~i=1,2,\dots
\end{equation}
This means
that ${\bf P}^*$ is the normalized eigenvector of the evolution matrix 
${\bf E}$ with an eigenvalue of $1$.
Once the steady state ${\bf P}^*$ is
calculated, it is possible to evaluate the steady-state average upward
growth probability,
\begin{equation}
\langle p_{\rm up} \rangle^*=\sum_{j=1}^{N_c}P^*_jp_{\rm up}(j),
\label{pup=}
\end{equation}
where $p_{\rm up}(j)$ is the total upward growth probability for configuration
$j$. The average steady-state density of the aggregate is then given by
\begin{equation}
\rho(N)=\frac{1}{N\langle p_{\rm up} \rangle^*}.
\label{rho(N)=1/Npup}
\end{equation}
Here, the density is written explicitly as a function of $N$. By $\rho(N)$ we
denote the true value of the density, in the limit $O \to \infty$.
We denote the
result of the $O$'th-order approximation by $\rho_c(N,O)$.

As mentioned, the number of configurations grows exponentially with $O$
and $N$, so it becomes infeasible to make the calculation for high values
of $O$ and $N$. We perform calculations for $N \leq 10$. The calculated
densities and the number of configurations are presented in Sec. 
\ref{EnumerationSec}. We find that in order
to obtain a relative accuracy of about $10^{-4}$, it is necessary to 
go up to $O=N-2$, or $O=N-1$. This is achieved for $N \leq 8$, but
for $N=9,10$ it is too heavy a task for our computer resources.
In spite of this, we are able to successfully extrapolate $O$ to infinity for
$N=9$ and $N=10$. 
The calculated densities are compared to direct measurements from cylindrical
site-DLA simulations, and are found to be the same up to
the accuracy of the simulation, which is about $0.01\%$ or better.

The fractal dimension $D$ is extracted from the assumption that 
$\rho(N) \propto N^{-(d-D)}$, where $d=2$ is the Euclidean dimension.
In general one should also expect some corrections to scaling, 
especially for low $N$'s, i.e.,
\begin{equation}
\rho(N)=AN^{-(d-D)}\left(1+B/N^{\theta}+\dots\right),  
\label{rho(N)=AN^(D-d)}
\end{equation}
where $A$ and $B$ are some constants,
$\theta$ represents the leading correction exponent,
and the dots stand for a series of higher
powers of $1/N$. This scaling hypothesis is validated by both the analytic
enumeration computation and by the simulations, 
which were conducted up to $N=128$. The best fit of such a model to the 
enumeration data
results in an estimate of the fractal dimension of cylindrical DLA in 2D,
$D=1.68\pm0.01$. 
The same fit is also performed with the simulation data, yielding 
$D=1.671\pm0.001$. 
This is to be compared with the value $D\approx1.66$ often found in
the literature \cite{Erzan95}.

The differences between bond-DLA and site-DLA are manifested in the 
boundary conditions for the Laplace equation, 
and in the way the growth probabilities are
extracted from the potential $\Phi$.
The boundary conditions at the top are,
\begin{equation}
\lim_{m\to\infty}{\partial \Phi(m,n)\over \partial m}
=1,~~n=0,\dots,N-1,
\end{equation}
where ${\bf \hat m}$ is the vertical direction (the growth direction), and
${\bf \hat n}$ denotes the periodic lateral direction. 
This describes a uniform flux of incoming
particles. In the original
DBM papers \cite{DBM84a,DBM84b}
a uniform potential is used instead of a uniform gradient,
but if the distant boundary is very far, then the differences between the 
solutions for the two cases is exponentially small \cite{Kol98}. 
The determination of the boundary
conditions on the aggregate should be done with care. In the case of bond-DLA
the potential is set to $0$ on the aggregate itself, while in site-DLA the 
potential should be set to $0$ on nearest neighbors sites, i.e., on sites 
where growth might occur. 
Also, the derivation of the growth probabilities from the potential in
site-DLA is done a bit differently than in bond-DLA,
as explained in the next section.

The Laplace equation with these boundary conditions can be solved exactly
\cite{Kol98,Kol2000a}. The idea is to divide the plane (or space in
higher dimensions) into two parts:
the upper part is an empty semi-infinite rectangle that begins at the row of 
the highest site on the lower boundary and continues upward ad infinitum. 
The lower boundary is the set of sites on which the potential $\Phi$ is 
set to $0$, and it depends on the type of sticking-conditions used:
In the case of bond-sticking conditions the boundary is the aggregate
itself, whereas in the case of site-sticking conditions it is the set
of sites that are nearest neighbors to the aggregate.
The lower part 
contains the aggregate and extends from the highest row downwards.
The row that contains the highest particle in the aggregate is usually set
as a reference row with $m=0$. Thus, in bond-DLA, 
the upper part has $m \geq 0$, and the lower part has $m \leq 0$, and
in site-DLA,
the upper part has $m \geq 1$, and the lower part has $m \leq 1$
, as explained in more detail in Sec. \ref{EnumerationSec}.
Note that in either case the dividing row is considered to belong to both
parts.
In the upper part, it is possible
to express the potentials in row $m+1$ as a linear combination of those
in row $m$,
\begin{equation}
\Phi(m+1,n)=1+\sum_{n'=0}^{N-1}\Phi(m,n')g_N(n-n').
\label{Phi(m+1,n)=1+sum}
\end{equation}
This is especially useful for the bottom row of the upper
part, $m=0$ or $m=1$ (depending on the type of sticking conditions).
The boundary Green's function $g_N(n)$, appearing in Eq. 
(\ref{Phi(m+1,n)=1+sum}), is given by
\begin{equation}
g_N(n)=\frac{1}{N}\sum_{l=0}^{N-1}e^{-\kappa_l}\cos(k_ln).
\label{g_N(n)=}
\end{equation}
The finite set of allowed wave-vectors $k_l={2\pi \over N}l$ for 
$l=0,\dots,N-1$, is imposed by the horizontal periodicity. The factor
$\kappa_l$ is related to $k_l$ through the dispersion relation
\begin{equation}
\sinh\left(\kappa/2\right)=\pm\sin\left(k/2\right),
\end{equation}
or more explicitly, 
\begin{equation}
e^{-\kappa(k)}=2-\cos(k)-\sqrt{\left(2-\cos(k)\right)^2-1}.
\end{equation}
An interesting property of the Green's function is that
\begin{equation}
\sum_{n=0}^{N-1}g_N(n)=1.
\label{sumg(n)=}
\end{equation}
This property was proved algebraically in Ref. \cite{Kol98} and was used
in Refs. \cite{Kol98,Kol2000a} to check the computations of the Green's
function. It is also used in the sample calculation of Sec.
\ref{SampleCalculationsSec} in the current paper for the same purpose.   

Usually, there is no general derivation for the solution in the lower part.
In spite of that, the number of sites in the lower part is finite and not
too large, so it is possible to simply write the equations for each of the 
potentials. The solution of the resulting finite and linear set of equations
is then straightforward. 

The paper is organized as follows: In Sec. \ref{EnumerationSec} we 
present in detail the differences in the computation of the growth 
probabilities between bond and site sticking conditions. This presentation also
explains the connection with the Laplace equation more rigorously. After that
we perform a few sample calculations, for $N=2$ and $N=3$, in order to 
demonstrate the method presented in the introduction. We then report the 
results of the computations for $N$ between $4$ and $10$ for various orders
of approximations $O$. 
We point out that the results collapse onto a universal function that enables
the extrapolation $O \to \infty$ for $N=8$, $9$ and $10$. 
This extrapolation
reduces the error appreciably. In Sec. \ref{SimulationSec} we present the 
simulation we made in order to verify our theoretical predictions.
This presentation also explains how
the boundary Green's function $g_N(n)$ is used in some way as a probability
function, in order to make the simulation more efficient. 
We summarize in Sec. \ref{SummarySec}.

\section{Enumeration}
\label{EnumerationSec}
Our computation method is referred to as enumeration, because it involves a
systematic processing of some complete lists of configurations. 
\subsection{The differences between site and bond sticking conditions} 
\begin{figure}
\epsfxsize 8cm
\epsfbox{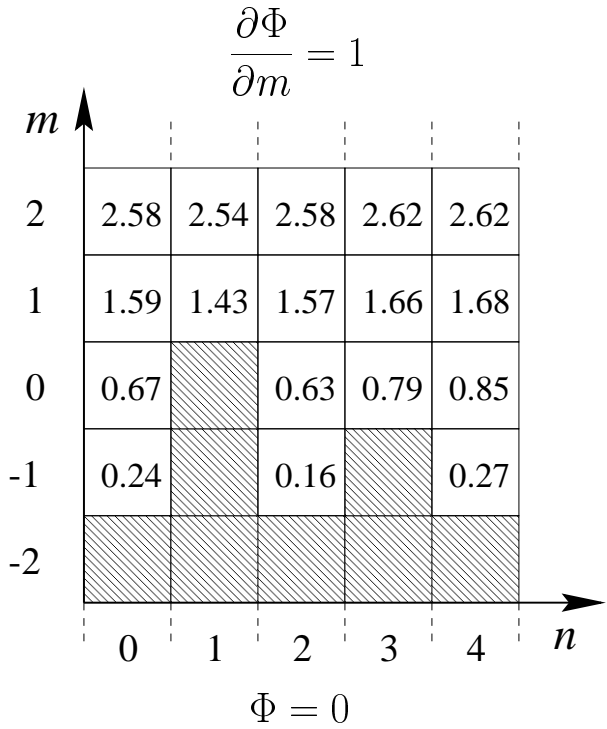}
\vspace{5mm}
\caption{An example of the solution of the Laplace equation 
$\nabla^2\Phi(m,n)=0$,
with boundary conditions $\Phi=0$ on the aggregate, and 
$\partial\Phi/\partial m=1$ on the upper distant boundary. Here, 
the width is $N=5$ and there
are periodic boundary conditions on the sides. The axes indicate the 
directions of the coordinates $m$ and $n$. These boundary conditions
are consistent with bond-sticking conditions.}
\label{DBMfig}
\end{figure}

\begin{figure}
\epsfxsize 8cm
\epsfbox{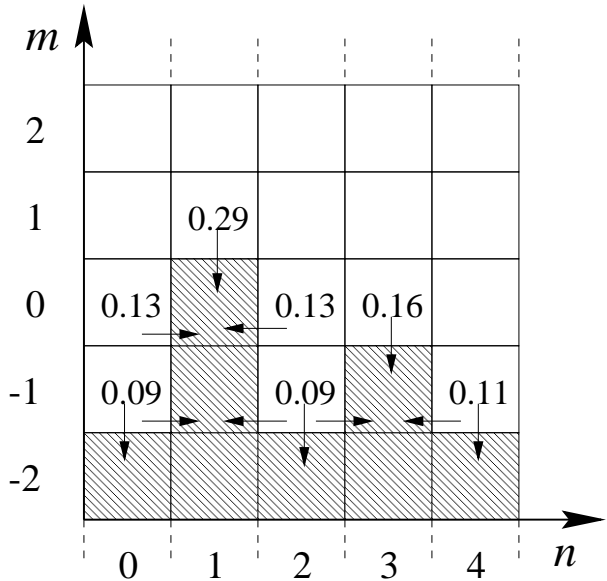}
\vspace{5mm}
\caption{The growth probabilities for the aggregate shown in Fig. 
\protect\ref{DBMfig}.
The growth probability in each perimeter site is proportional
to the potential $\Phi$ at that site and to the number of bonds $N_b$ leading
from the site into the aggregate (denoted by arrows), e.g., $N_b=3$ for the
site at $(-1,2)$ and $N_b=1$ for the site $(0,2)$, right above it.}
\label{BondMultiplicityFig}
\end{figure}

Before proceeding with the actual calculations, we point out the differences
in the computation of the growth probabilities between bond and site sticking
conditions, because these differences are the essence of this paper. 
In order to 
point out the differences, we first review the method for computing the
growth probabilities with bond-sticking conditions. The first step
is to solve the Laplace equation (\ref{Laplace}) on a lattice,
where the boundary conditions are $\Phi=0$ on the aggregate, and 
$\partial\Phi/\partial m=1$ on the distant boundary. 
In the case of cylindrical geometry there are periodic boundary conditions on 
the sides, see Fig. \ref{DBMfig}. The sticking probability per bond is then
given by
\begin{equation}
p_b={\Phi_b\over\sum_{b'}\Phi_{b'}},
\end{equation}
where the subscript $b$ refers to a perimeter bond and $\Phi_b$ 
refers to the potential
difference across such a bond. Because the potential is null on the
aggregate, this difference is equal to the value of the potential in a
nearest neighbor site. Finally, the growth probability 
per site is computed by multiplying the sticking probability per bond
by the number of bonds associated with the site, $N_b$, see Fig. 
\ref{BondMultiplicityFig}.

Why does this procedure give the exact growth probabilities? In order
to answer this question we must return to the original definition of DLA, 
that involves a single random walker. The random walker is injected into
a random site near the remote boundary and it diffuses until it attempts
to penetrate the aggregate, in which case it gets stuck. By ``penetrate''
we mean that it is not sufficient for the random walker to be in a nearest
neighbor site to the aggregate, but that it has to attempt to cross a 
perimeter bond in order for it to stick. A possible way of measuring the
growth probabilities for a particular interface configuration is to send many
random walkers, one after the other, and remove them after they stick. One
has to keep track of how many particles get stuck in each site. Eventually,
the growth probabilities per site are estimated by the fraction of particles
that got stuck in each site. Instead of releasing the random walkers
one at a time, it is more efficient to release many of them simultaneously,
and let them perform a random walk without interacting with each other.
Moreover, instead of releasing a large amount of particles in a single batch
and waiting until all of them stick, it is also possible to inject them at a 
constant rate near the boundary,
i.e., in each time step inject a new particle into each site near the boundary
with a uniform probability $r$. 
The advantage in this way of performing the measurement
is that after an initial equilibration time the system arrives at a steady
state, which is characterized by a time-independent average number of
random walkers in all of the sites, including sites that are not near any of
the boundaries. In the steady state the average number of random walkers
entering into the system in each time step at the upper boundary is equal 
to the average number of random walkers vanishing out of the lower boundary. 

Denote the average number of random walkers in each site 
in the steady-state by $\Phi(m,n)$.
The crucial point is that $\Phi$ is time independent. 
In order to calculate
$\Phi$ we note that it satisfies the discrete Laplace equation
\begin{eqnarray}
&0=\nabla^2\Phi(m,n)\equiv-4\Phi(m,n)+& \\ \nonumber
&\Phi(m+1,n)+\Phi(m-1,n)+\Phi(m,n+1)+\Phi(m,n-1)&,
\end{eqnarray}
because every random walker is equally probable to go to any one of its 
nearest neighbor sites.
Thus, on a general lattice (or graph) the Laplace equation states that 
the value of $\Phi$ at each
site is equal to the mean value of $\Phi$ on its nearest neighbor sites. 
Special 
care should be give to sites near the boundaries. Near the upper boundary each
site has only three nearest neighbors, and particles are added at a constant
rate $r$, therefore,
\begin{eqnarray}
&\Phi(m,n)=\frac{1}{4}\left[\Phi(m,n-1)+\Phi(m,n+1)\right.& \nonumber \\
&\left.+\Phi(m-1,n)+\Phi(m,n)\right]+r&.
\label{TopBoundaryConditionsEq}
\end{eqnarray}
Note that the last term on the left before $r$ is $\Phi(m,n)$, instead of
$\Phi(m+1,n)$, because the particles that randomly choose to go up
are unable to do so because of the boundary, and therefore they remain in the
same place. Now, let us define $\Phi(m+1,n)\equiv\Phi(m,n)+4r$ as a fictitious
density above the boundary. Then we see that Eq. 
(\ref{TopBoundaryConditionsEq}) turns into the standard Laplace equation. 
This shows that instead of using the injection rate parameter $r$, 
it is possible to use
the regular Laplace equation with the Neumann type boundary conditions that 
require the specification of the electric field, which
corresponds to the difference in the potential across the upper boundary.
Since the value of $r$ does not change the growth 
probabilities, we are free to choose any value for it. 
If, for example, we choose $r=1/4$ then the boundary conditions at the top are
\begin{equation}
\frac{\partial \Phi}{\partial m}\equiv\Phi(m+1,n)-\Phi(m,n)=1,
\end{equation}
for $n=0,1,\dots,N-1$. We choose the upper boundary to be very far away from
the lower one, 
because this simplifies the analytic expressions involved in the solution
of the Laplace equation, while leaving the sticking probabilities 
practically unchanged.

Near the bottom boundary the situation is a bit different. Each random
walker that attempts to go into the aggregate is taken out of the system.
The steady state equation for these sites is therefore
\begin{equation}
\Phi(m,n)=\frac{1}{4}\sum_{\rm nn}\Phi(m',n'),
\end{equation}
where the sum is taken over all the sites 
$(m',n')$ that are nearest neighbors (nn) to $(m,n)$.
At the lower boundary the situation is similar, because once 
again, we obtain the regular Laplace equation, if we choose the boundary
conditions $\Phi=0$ on the aggregate itself.

The growth probability in each site is evaluated as the average number
of random walkers that stick in that site per unit time, normalized by the 
total
number of particles sticking in a time step across the total length of the 
lower boundary. An average of $1/4$ of the particles in a site choose
to go in each direction. Particularly, a fraction of $1/4$ of the particles
vanish after choosing to go via bonds that connect to the aggregate. The 
average total number of particles sticking in a site would be a sum over all
of its interface bonds, $\sum_b\Phi/4=N_b\Phi/4$. In the steady-state,
the average total number of sticking random walkers is equal to the average
total number of random walkers injected into the system. Near the
upper boundary $r=1/4$ random walkers are injected into each of the
$N$ sites. Therefore the normalization factor is $N/4$ and the growth 
probability in each site is $N_b\Phi/N$. In Ref. \cite{Kol2000a} we arrive
at the same result using the discrete Gauss theorem.

\begin{figure}
\epsfxsize 8cm
\epsfbox{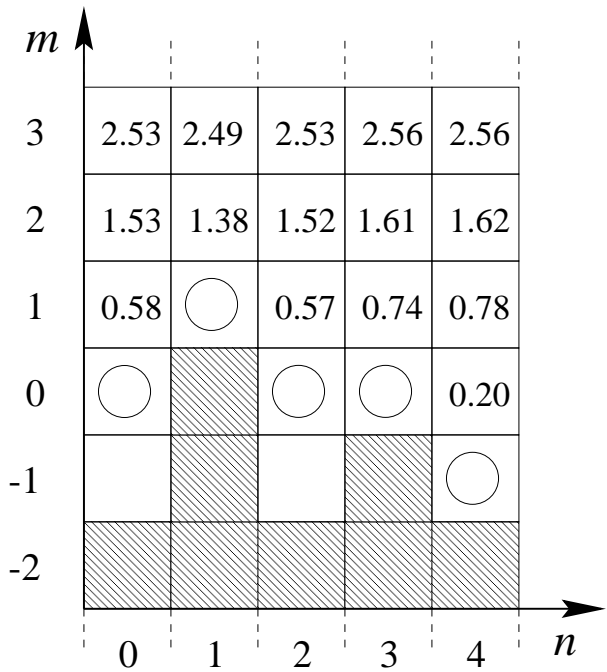}
\vspace{5mm}
\caption{The solution to the Laplace equation near the same aggregate as
in Figs. \protect\ref{DBMfig} and \protect\ref{BondMultiplicityFig}, 
only with site-sticking
conditions. The circles denote the perimeter sites where a random walker
might stick. The boundary conditions are that $\Phi=0$ on these sites,
unlike the case of bond-sticking conditions where $\Phi=0$
on the aggregate itself. The boundary conditions $\partial \Phi/\partial m=1$
at large $m$ remain unchanged.}
\label{SitePotentialsFig}
\end{figure}

\begin{figure}
\epsfbox{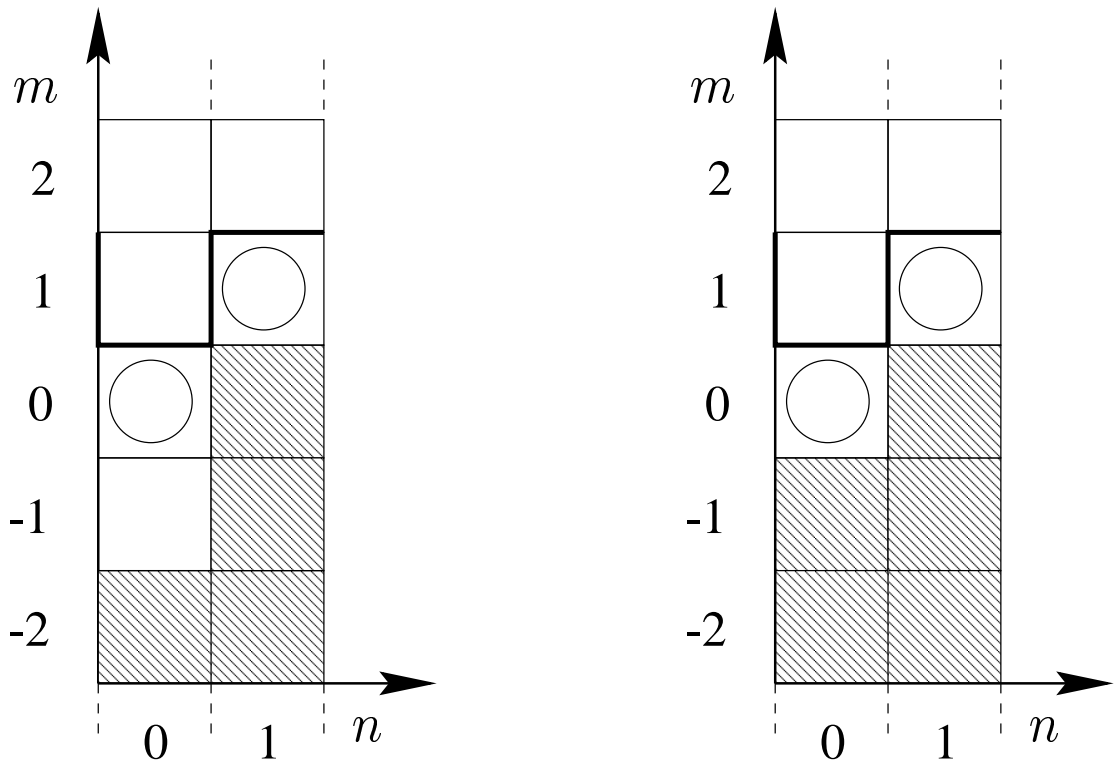}
\vspace{5mm}
\caption{Two different aggregates (represented by the dashed squares)
with $N=2$, having
exactly the same set of sites where a random walker may stick (shown in
circles), and thus having the same boundary (the bold line) for the Laplace 
equation, where $\Phi=0$.}
\label{BoundariesFig}
\end{figure}

The situation in site-DLA is different in the boundary conditions used
near the aggregate, and in the expression for the growth probabilities.
Now, random walkers never arrive at sites that are nearest neighbors to the
aggregate, because as soon as they do they get stuck and are immediately
removed from the system. We therefore impose $\Phi=0$ not on the aggregate 
itself, but rather on its nearest neighbor sites, see Fig.
\ref{SitePotentialsFig}. In general, the boundary for the Laplace equation
is obtained by coating the aggregate with a layer of circled sites, as
shown in the figure.
This differentiates between the boundary
of the aggregate itself and the boundary for the Laplace equation, 
in the sense that it is possible that two different aggregates would have the
same boundary for the Laplace equation, see Fig. \ref{BoundariesFig}.
One can think that a random walker does not interact directly with the 
boundary of the aggregate, but rather, it interacts with the circled sites that
make the boundary for the Laplace equation. This means that
a random walker that obeys site-sticking conditions cannot be used as a probe
in any way to determine which of the two aggregates in the figure are present. 
Consequently, any two different aggregates that have the same boundary for the 
Laplace equation must have the exact same set of growth probabilities and 
can be therefore considered as equivalent.
Thus, from now on when we refer
to an interface configuration, we relate to the shape of the boundary for
the Laplace equation. 
The probability to be in such a configuration is a 
sum over all the underlying aggregate configurations. Another effect
of the transition to the Laplace boundary is the narrowing of fjords. The
padding of the aggregate by circled sites causes all the fjords
to be narrower by two sites. Thus, a random walker can only penetrate
into aggregates that have branches that are at least three sites apart.

\begin{figure}
\epsfxsize 8cm
\epsfbox{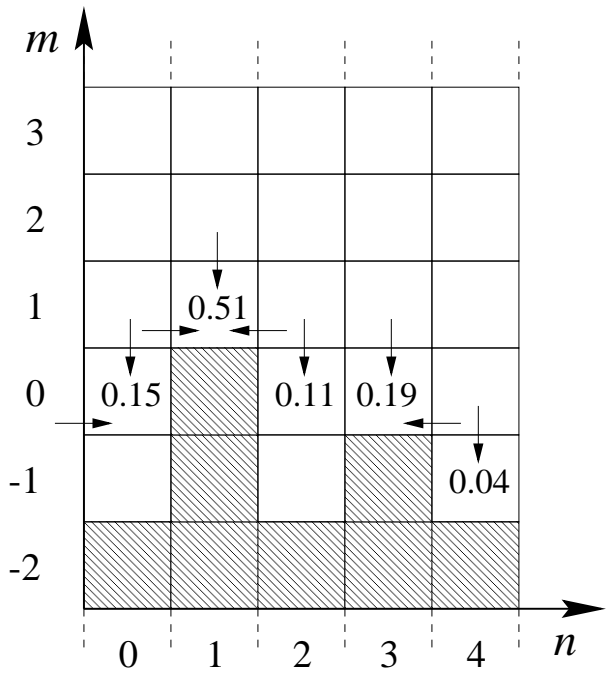}
\vspace{5mm}
\caption{The sticking probabilities in each of the circled sites of
Fig. \protect\ref{SitePotentialsFig}. They are computed by summing
over all of the bonds that go into each site (denoted by arrows), unlike
the case of bond-sticking conditions, where contributions are summed over
bonds that go out of each site .}
\label{SiteGrowthProbabilitiesFig}
\end{figure}

As in the case of bond-DLA, the sticking probabilities are evaluated as 
fractions of the average
number of random walkers that stick per unit time. Only now, we must sum
over bonds that lead {\it into} the site, rather than out of it, as is
the case in bond-DLA. The growth probability per site is therefore,
\begin{equation}
p_{\rm site}(m,n)=\frac{1}{N}\sum_{\rm nn}\Phi(m',n'),
\end{equation}
where $p_{\rm site}(m,n)$ is the total sticking probability at the perimeter 
site $(m,n)$,
see Fig. \ref{SiteGrowthProbabilitiesFig}. Unlike the case of bond-sticking
conditions, where a single potential determines the sticking probability in
a particular site, now the potentials in several different sites contribute.
This is because in bond-DLA the random walker sticks before it
moves out of the site, whereas in site-DLA the random walker sticks after
it moves into it. This difference gives the upper most tip of the
aggregate even a greater advantage relative to bond-DLA, because a 
single particle tip gathers contributions from three sides in site-DLA, 
whereas in bond-DLA the only contribution is from above. This comes in
addition to the screening property of the Laplace equation
(common to both types of sticking conditions), which causes
the sticking probabilities at the lower parts of the interface to decrease 
exponentially. 
\subsection{Exact solutions for $N=2$, $3$}
\label{SampleCalculationsSec}
The best way of explaining the enumeration method 
is by showing some sample calculations
in detail. We present here the two simplest cases, namely, $N=2$ and $N=3$.
In these relatively simple cases there is only a finite number of 
configurations, so it is
possible to get an exact solution with no need for approximations.

\begin{figure}
\epsfbox{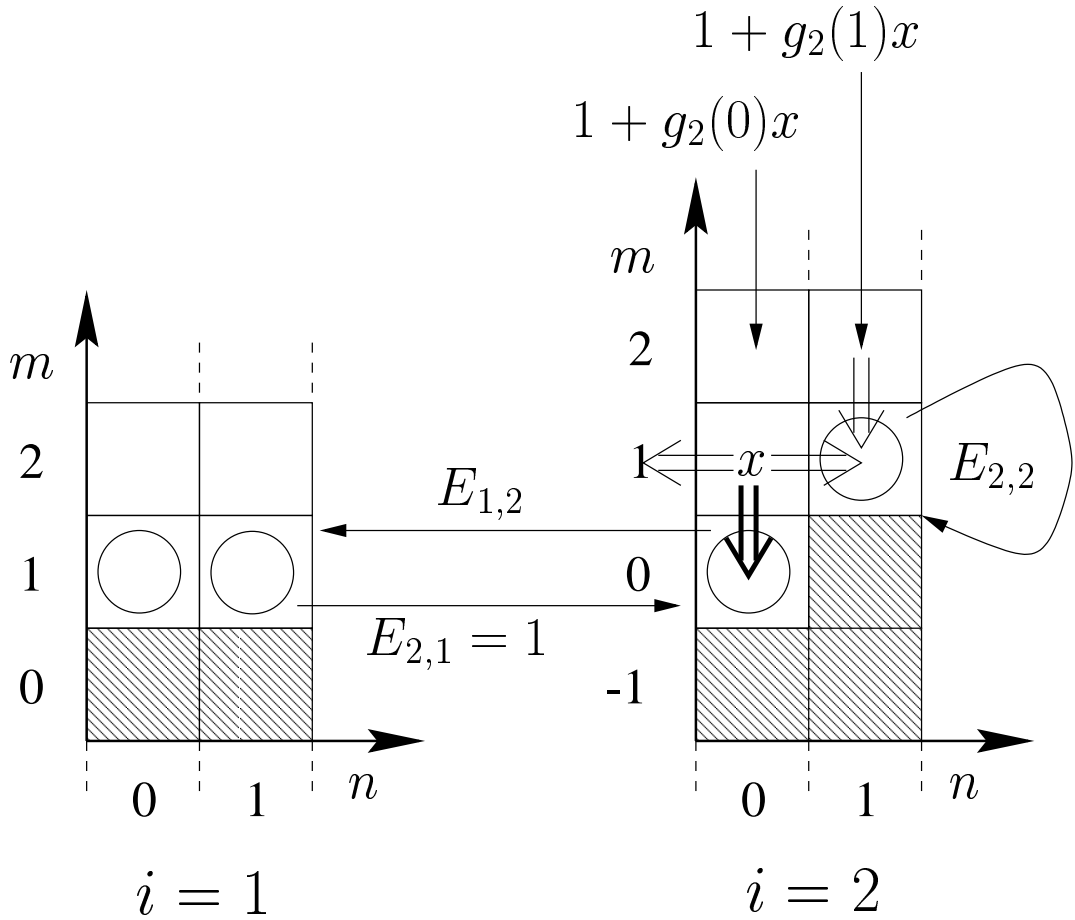}
\vspace{5mm}
\caption{The two possible configurations for $N=2$. The circles denote
the sites where a random walker might stick.
Also shown are the 
possible transitions between them, denoted by arrows, and the relevant
matrix elements $E_{i,j}$. The distribution of the potential $\Phi$ over the 
lattice is demonstrated only for configuration $i=2$, see text
for explanation. The double arrows ($\Downarrow$, $\Rightarrow$ and
$\Leftarrow$) show the bonds of the possible
access paths, which a random walker can take into the circled sites. The bold
double arrow shows the only bond going into site $(0,0)$. The other
three bonds lead into site $(1,1)$.}
\label{N2IndexFig}
\end{figure}

For $N=2$, the interface of the aggregate itself has an infinite number of 
possible configurations, because
it has the shape of a step whose height $j$ can be any nonnegative integer
\cite{Kol98}. However, in site-DLA there are only two distinct states: $j=0$
and $j>0$. The case $j=0$ refers to a flat interface, i.e., the two columns
have the same height, and a growth process will create a step with $j=1$,
with probability $1$. For any step size $j>0$ there are only two sites where
a random walker may stick: above the highest particle in the aggregate,
or on its side, see Fig. \ref{BoundariesFig}. 
There is no possibility for the random walker to penetrate
into a fjord in $N=2$, because it is too narrow, and the particle would stick
at its entrance. 
The two configurations are indexed by $i=1$ and $i=2$ 
respectively and are shown in Fig. \ref{N2IndexFig}. 

We now begin building the evolution matrix ${\bf E}$, by finding the growth
probabilities for each of the two configurations. As mentioned, configuration
$i=1$ turns into $i=2$ with probability $1$, hence $E_{1,1}=0$ and
$E_{2,1}=1$. It is important to keep track of the total upward
growth probability for each configuration, $p_{\rm up}(i)$, that
corresponds to events in which a newly stuck particle is higher than all
of the particles in the aggregate. In this case $p_{\rm up}(1)=1$.

In order to solve for $i=2$, we first have to compute the Green's
function according to Eq. (\ref{g_N(n)=}), which gives
\begin{eqnarray}
g_2(0)&=&2-\sqrt{2}=0.5858, \nonumber \\
g_2(1)&=&\sqrt{2}-1=0.4142.
\end{eqnarray}
We check our calculations by verifying that $g_2(0)+g_2(1)=1$, as expected
from Eq. (\ref{sumg(n)=}). The potential $\Phi$ near the
growth sites can be expressed in terms of the variable
$x\equiv \Phi(1,0)$ according to Eq. (\ref{Phi(m+1,n)=1+sum}), 
as shown in Fig. \ref{N2IndexFig}. 
We usually set the row containing the highest particle in the aggregate
as the reference row, with $m=0$. Thus, the row $m=1$ always contains the
highest circled site that belongs to the Laplace boundary.
Each of the sites in row $m=1$ contributes to the potentials in the sites in
row $m=2$. The weight of the contribution is equal to the value
of the Green's function $g_N(n)$, where $n$ is the horizontal distance between
the contributing site in row $m=1$ and the evaluated site in row $m=2$.
In this simple case there is only one site with a nonzero potential, namely,
$\Phi(1,0)$, which is yet unknown, and which we denote by $x$. The site
$(1,1)$ on its side is nearest neighbor to the aggregate and therefore we
set $\Phi(1,1)=0$. Thus, the potential of the sites in row $m=2$ have
only a  contribution from $x$. More specifically, $\Phi(2,0)=1+g_2(0)x$
because it is right above $x$, and $\Phi(2,1)=1+g_2(1)x$ because it is
removed by one site. 
The potential $\Phi(2,0)$ does not contribute
to any growth process, but is important for solving for $x$. The variable
$x$ is found using its Laplace
equation,
\begin{eqnarray}
4x&=&1+g_2(0)x, \nonumber \\
\Rightarrow x&=&\frac{2-\sqrt{2}}{2}=0.2929.
\end{eqnarray}
Growth in site $(0,0)$ results in the flat configuration $i=1$. It
can only occur via one bond from site $(1,0)$, denoted by a bold double
arrow (${\bf \Downarrow}$) in Fig. \ref{N2IndexFig}. Hence,
\begin{equation}
E_{1,2}=\frac{x}{2}=\frac{2-\sqrt{2}}{4}=0.1464,
\end{equation}
where the denominator comes from the normalization factor $N=2$.
Growth can also occur in site $(1,1)$. 
This time there are three different bonds
coming from two sites: there are two bonds coming from $(1,0)$ and an 
additional one coming from $(2,1)$. 
This upward growth results in the same configuration, so
\begin{equation}
E_{2,2}=p_{\rm up}(2)=\frac{1}{2}\left[2x+1+g_2(1)x\right]=\frac{2+\sqrt{2}}
{4}=0.8536.
\end{equation}

This concludes the calculation of all of the growth processes. The resulting
evolution matrix is
\begin{equation}
{\bf E}=\left[
\begin{array}{cc}
0~&0.1464 \\
1~&0.8536
\end{array}
\right].
\end{equation} 
We verify that the matrix is properly normalized by noting that the sum of
the terms in each of its columns is equal to $1$, i.e., 
\begin{equation}
\sum_{i=1}^{2}E_{i,j}=1,~~j=1,2.
\label{MatrixNormalizationEq}
\end{equation}
The general theorem mentioned in the introduction 
ensures the existence of a single eigenvector with an
eigenvalue of $1$, or in other words, a fixed point vector ${\bf P}^*$ that
satisfies,
\begin{equation}
{\bf P}^*={\bf EP}^*.
\end{equation}
The fact that the process
is closed is manifested in Eq. (\ref{MatrixNormalizationEq}). The
process is also irreducible because there is a finite probability to go
from any initial state to any final state during a finite number of time steps.
The fact that there is a single fixed point implies that starting from any 
initial state, the system will converge to the fixed point. This fixed point
represents the asymptotic time probabilities for seeing
either one of the two possible configurations.

The eigenvalues are the roots of the characteristic polynomial,
\begin{eqnarray}
&&\lambda_0=1, \nonumber \\
&&\lambda_1=-\frac{2-\sqrt{2}}{4}=-0.1464,
\end{eqnarray}
and the normalized fixed-point vector ${\bf P}^*$ is given by
\begin{equation}
\begin{array}{c}
P_1^*=\frac{5-\sqrt{8}}{17}=0.1277,\\ \nonumber
P_2^*=\frac{12+\sqrt{8}}{17}=0.8723.
\end{array}
\end{equation}
The steady-state weights enable us to calculate the average 
upward growth probability,
\begin{equation}
\left<p_{\rm up}\right>^*=\sum_{i=1}^2P^*_ip_{\rm up}(i)=
\frac{12+\sqrt{8}}{17}=0.8723,
\end{equation}
which is connected to the mean density,
\begin{equation}
\rho(2)={1 \over 2\left<p_{\rm up}\right>^*}={6-\sqrt{2} \over 8}=0.5732.
\end{equation}

It is also possible to calculate the rate of convergence to the steady state.
In general, the rate of convergence is determined by the largest eigenvalue
of ${\bf E}$, other than $1$. Suppose that at time $t=0$ the state of the
system differs from the steady state ${\bf P}(0) \neq {\bf P}^*$.
The difference vector 
\begin{equation}
{\bf v}(0)\equiv{\bf P}(0)-{\bf P}^*,
\end{equation}
belongs to the linear subspace of vectors $V=\{{\bf v}|\sum v_i=0\}$, because
both ${\bf P}(0)$ and ${\bf P}^*$ are normalized probability vectors and
thus the sum of their components is equal to $1$. Now, Eq. 
(\ref{MatrixNormalizationEq}) ensures that $V$ is an eigen-subspace of
${\bf E}$, and as such it must contain at least one eigenvector. Since
in this simple case
the space of configurations is only two-dimensional, then
$V$ is one-dimensional, and ${\bf v}(0)$ 
is necessarily an eigenvector 
with the eigenvalue $\lambda_1=-0.1464$. 
After $t$ time
steps the state of the system is
\begin{equation}
{\bf P}(t)={\bf E}^t{\bf P}(0)={\bf P}^*+\lambda_1^t{\bf v}(0).
\end{equation}
Therefore, the deviation from the steady state decays exponentially,
\begin{equation}
{\bf P}(t)-{\bf P}^*=\lambda_1^t{\bf v}(0)=
(-1)^te^{-\frac{t}{\tau}}{\bf v}(0),
\end{equation}
where
\begin{equation}
\tau \equiv -{1 \over \log|\lambda_1|}=0.5584.
\end{equation}

This means that a single time step is practically sufficient to arrive at the
steady state.
All of these theoretical predictions agree with results obtained from numeric 
simulations, up to the accuracy of the simulation, which is better
than $10^{-5}$.
This dynamics is actually exactly the same as the first-order approximation of
the frustrated climber model in Ref. \cite{Kol2000a}, except that the analysis
of the temporal convergence is a little bit more refined there.

\begin{figure}
\epsfbox{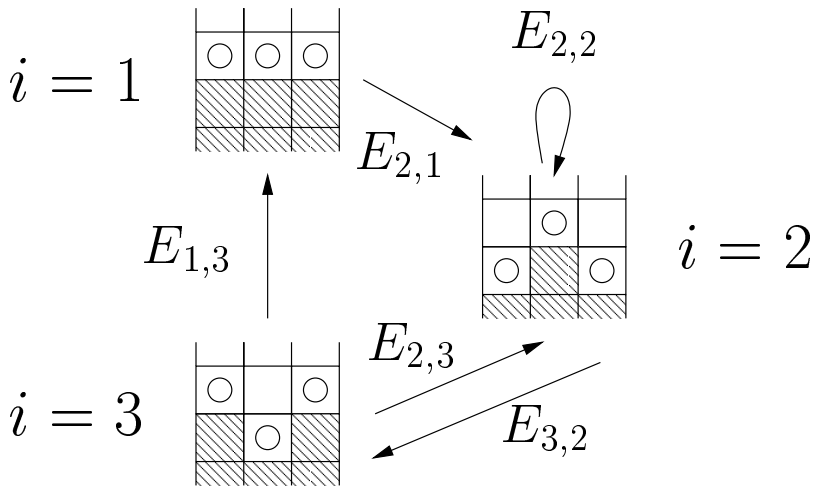}
\vspace{5mm}
\caption{The possible configurations and the possible transitions between them
for $N=3$.}
\label{N3IndexFig}
\end{figure}

The solution of the case $N=3$ is also relatively simple, because again there
is only a finite number of growth configurations. This is because
the width of the widest possible fjord is two sites, which is still 
insufficient for
a random walker to penetrate, i.e., a 
random walker sticks as soon as it enters into a
fjord. The three possible configurations are indexed in Fig. \ref{N3IndexFig}. 
These are the same as the three configurations of the first-order approximation
for bond-DLA with $N=3$ \cite{Kol2000a}.
As in the example of $N=2$, we proceed to calculate the 
probabilities for every growth process in each of the configurations. 
Once again, we first calculate the Green's function,
\begin{equation}
\begin{array}{c}
g_3(0)={6-\sqrt{21}\over 3}=0.4725, \\
g_3(1)=g_3(2)={1-g_3(0)\over 2}={\sqrt{21}-3\over 6}=0.2638.
\end{array}
\end{equation}

\begin{figure}
\epsfbox{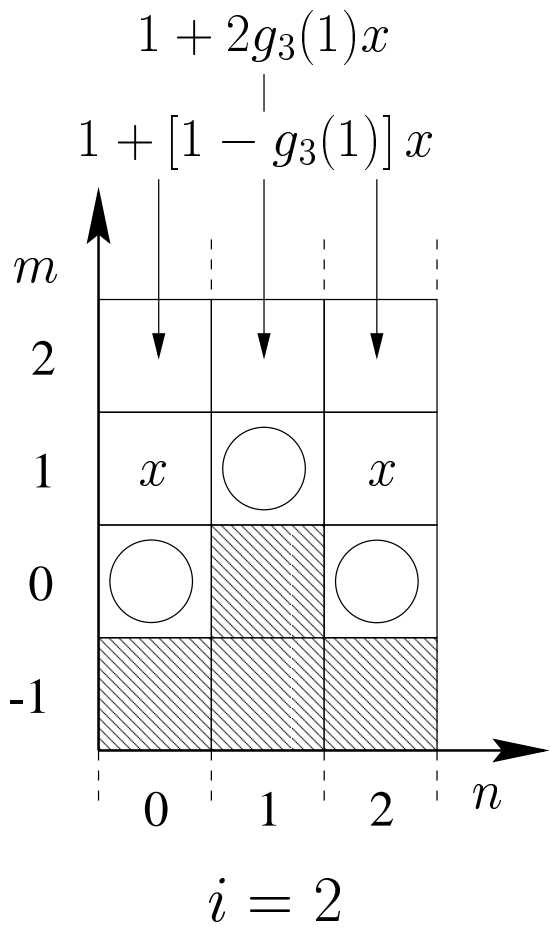}
\vspace{5mm}
\caption{A ``potential diagram'': the potentials $\Phi(m,n)$ of configuration
$i=2$, expressed in terms of the variable $x$.}
\label{N3i=2Fig}
\end{figure}

The first configuration, $i=1$, grows with probability $1$ into configuration
$i=2$. Thus, $E_{2,1}=1$ and $E_{1,1}=E_{3,1}=0$, and also $p_{\rm up}(1)=1$.
The potential diagram for $i=2$ is shown in Fig. \ref{N3i=2Fig}.
Because of symmetry
it is possible to conclude that $\Phi(1,0)=\Phi(1,2)=x$. The Laplace equation
for $x$ is
\begin{eqnarray}
4x&=&x+1+\left[g_3(0)+g_3(1)\right]x, \nonumber \\
\Rightarrow x&=&{9-\sqrt{21}\over 10}=0.4417.
\end{eqnarray}
The sticking probability at $(0,0)$ is $x/3$, because there is a single 
connecting bond, and because the normalization factor is $1/3$ for this case.
The resulting configuration is $i=3$,
however, a sticking event at $(0,2)$ also leads to $i=3$, so that
$E_{3,2}=\frac{2}{3}x=\frac{9-\sqrt{21}}{15}=0.2945$. 
The other possibility is an upward
growth at $(2,1)$, that results in the initial configuration $i=2$. Thus,
$E_{2,2}=p_{\rm up}(2)=1-E_{3,2}=\frac{6+\sqrt{21}}{15}=0.7055$, 
and $E_{1,2}=0$.

\begin{figure}
\epsfbox{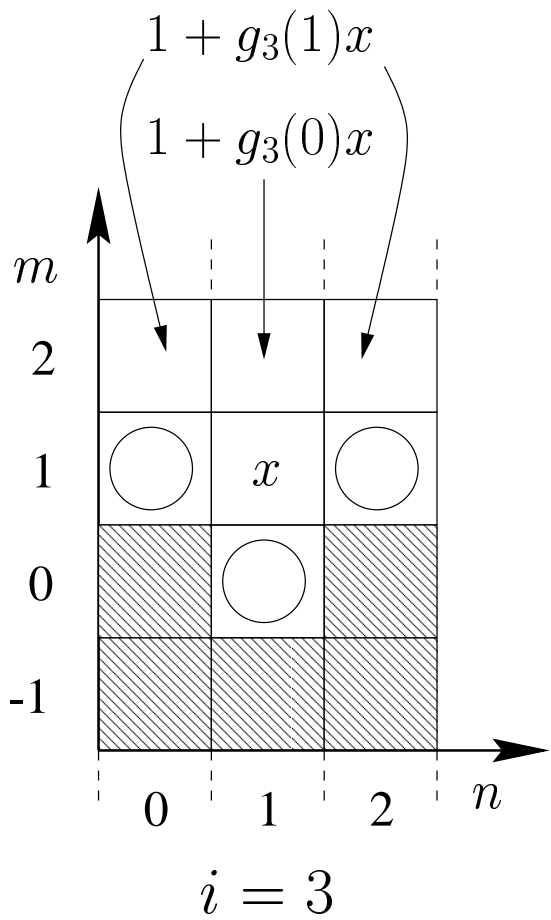}
\vspace{5mm}
\caption{The potential diagram for configuration $i=3$.}
\label{N3i=3Fig}
\end{figure}

The potential diagram for $i=3$ is shown in Fig. \ref{N3i=3Fig}. 
The Laplace equation is
\begin{eqnarray}
4x&=&1+g_3(0)x, \nonumber \\
\Rightarrow x&=&{6-\sqrt{21}\over5}=0.2835.
\end{eqnarray}
A sticking event in $(0,1)$ leads to $i=1$, therefore $E_{1,3}=x/3$.
The other possible sticking events at $(1,0)$ or $(1,2)$ involve upward 
growths, that result in $i=2$, i.e., $E_{2,3}=p_{\rm up}(3)=
1-x/3=\frac{9+\sqrt{21}}{15}=0.9055$. This completes the calculation
of all of the element of the evolution matrix:
\begin{equation}
{\bf E}=\left[
\begin{array}{ccc}
0&0& 0.0945 \\
1&0.7055&0.9055 \\
0&0.2945&0
\end{array}
\right].
\end{equation}
The normalized fixed point of the matrix is
\begin{equation}
{\bf P}^*=\left[
\begin{array}{ccc}
0.0210,&0.7562,&0.2227
\end{array}
\right].
\end{equation}
This enables the computation of the average upward growth probability, and of 
the average density:
\begin{eqnarray}
&\left<p_{\rm up}\right>^*=\sum_{j=1}^3P^*_jp_{\rm up}(j)=0.756245,& 
\nonumber \\
&\rho=\frac{1}{3\left<p_{\rm up}\right>^*}=0.440774.&
\end{eqnarray}
The second largest eigenvalue determines the characteristic time constant of
the exponential convergence to the steady state,
\begin{equation}
\tau=-{1\over\log|\lambda_1|}=0.56.
\end{equation}
\subsection{Approximations for $N>3$} 
\begin{figure}
\epsfbox{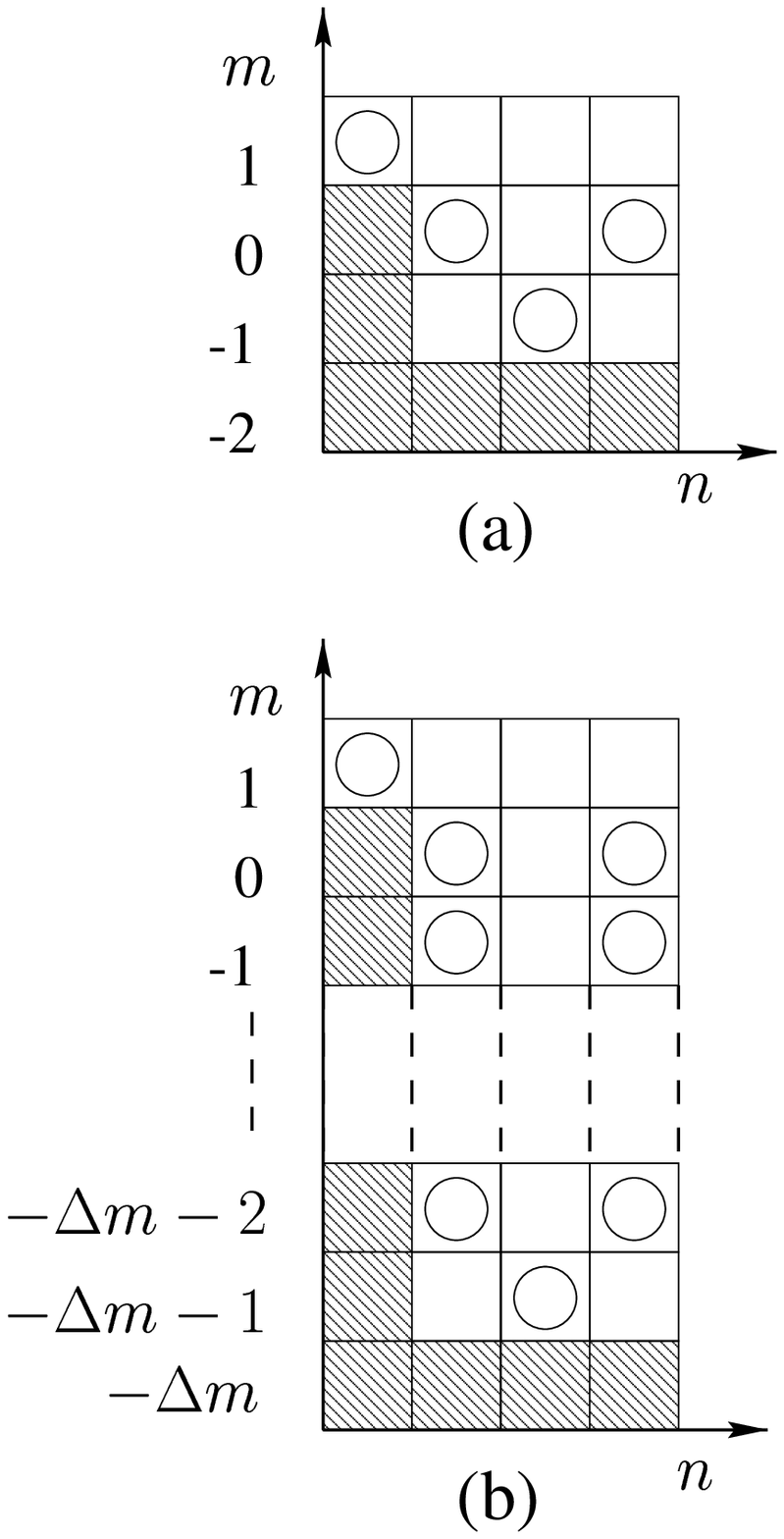}
\vspace{5mm}
\caption{The only configurations for a cylinder of width
$N=4$: (a) $\Delta m=2$, (b) $\Delta m>2$.}
\label{N4ExampleFig}
\end{figure}

\begin{figure}
\epsfbox{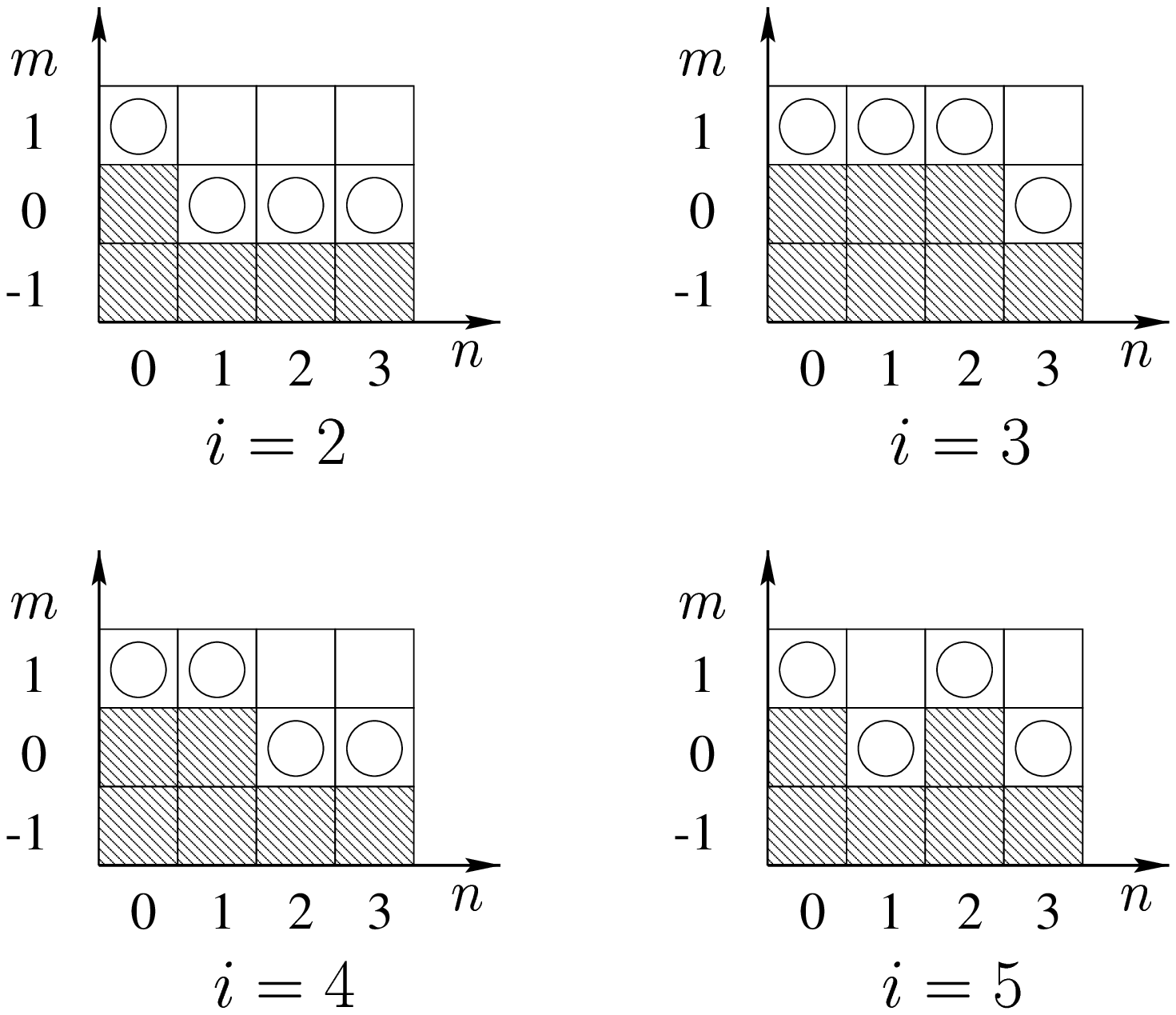}
\vspace{5mm}
\caption{The four possible configurations for $N=4$ with $\Delta m=1$.
These configurations are indexed between $i=2$ and $i=5$, and
the flat configuration with $\Delta m=0$ is indexed by $i=1$.}
\label{N4Dm=1Fig}
\end{figure}

The two examples of the previous sections, for $N=2$ and $N=3$, are
special because there is only a finite number
of possible configurations; a random walker cannot enter
a fjord whose width is less than three sites when using site sticking 
conditions.
The case $N=4$ is the narrowest cylinder that can have a fjord 
that is three sites wide. Since this fjord can be arbitrarily deep, there is an
infinite number of configurations. In spite of that, every configuration
that has a fjord, which is more than one site deep, is uniquely determined
by its depth, i.e., there is only one configuration with $\Delta m=2$,
a single configuration with $\Delta m=3$, and in general:
a single configuration with a specific $\Delta m$, if $\Delta m\geq 2$.  
The unique configuration with $\Delta m=2$ is
shown in Fig. \ref{N4ExampleFig}, along with 
the single configuration with a specific  $\Delta m$ that is larger than two.
Other than that, there are four possible configurations with $\Delta m=1$, 
which are shown in Fig. \ref{N4Dm=1Fig}, and finally, the 
trivial flat configuration, with $\Delta m=0$.

This case resembles bond-DLA with 
$N=2$ \cite{Kol98}, in the sense that in both cases there is an infinite number
of configurations, but this infinity can be represented using a single 
parameter. In Ref. \cite{Kol98} this parameter is called ``the step size''
and is denoted by $j$, but actually it is the same as $\Delta m$. The
case of site-DLA with $N=4$ is a bit different, because there are four
configurations with $\Delta m=1$ instead of one. There is also a resemblance
between the solution of the Laplace equation for the two cases, because
in both cases the Laplace equation is solved on a single column with
zero boundary conditions on the sides. Thus, in both cases there is an
exponential decay of the potential inside the fjord, which is governed by
the multiplicative factor $e^{-\kappa_f}=2-\sqrt{3}$. This enables us to
treat the current case in an analogous way to the previous one. This could
have given us analytic expressions for the Markovian matrix $E_{i,j}$,
$i,j=1,2,\dots,\infty$, for the steady-state vector $P^*_i$, 
$i=1,2,\dots,\infty$, and for the distribution of gaps inside
the aggregate. However, we omit the presentation of this calculation 
because it is not of main interest of this work, and so we treat the
case of $N=4$ in the same way as $N>4$. 

For $N>4$ the boundary may be complex, and it cannot be easily characterized
because the width of a fjord can fluctuate and overhangs may appear. 
We therefore use the approximation scheme described in the introduction,
which was also used for bond-DLA \cite{Kol2000a}. 
The calculation procedure involves going over all the possible 
configurations to some order, and calculating their set of growth 
probabilities. 
It is feasible to perform this task manually when the number
of configurations is relatively small, but as $N$ and $O$ increase, the number
of configurations grows exponentially and it becomes impractical to do so.
We use the same computer program that was used for 
bond-sticking conditions, after making the necessary adjustments due to the
site-sticking conditions. Manual calculations may still be
important as test cases
to check the operation of the program. 

The program goes over all of the possible configurations systematically. 
It starts with the trivial flat configuration ($\Delta m=0$), 
which is indexed by $j=1$.
This configuration has only one possible growth process, 
which occurs with probability $1$, that turns the 
interface into configuration $j=2$, which has a single bump.
The program then continues to $j=2$ and
analyzes its growth probabilities. Every growth process changes the shape of 
the boundary. Each time a particle sticks in a certain site, the program
has to identify the newly formed configuration. In order to do so, it 
marks all of the nearest neighbors of the newly attached particle, 
because new particles may stick there.
The new configuration is searched for in the 
existing list of configurations, which were already analyzed by the program.
If it does not exist then it is added at the end of the list. 
In either case the program identifies the index of the resultant configuration
$i$. Now, if the index
of the original configuration is $j$, then the growth probability
is stored in the matrix element $E_{i,j}$. 

A configuration is characterized using the set of sites that are connected
to infinity because these are the sites that are accessible to the random
walker. Of course, any site that is higher than the highest site on a
certain boundary is connected to infinity. Hence, it is sufficient to
specify only the set of sites that are not higher than the highest
site (the region $m \leq 1$).
A single growth process may cause a whole region of sites to 
disconnect from infinity, for instance by sealing off an entrance to a 
fjord. This means that it is not sufficient to mark the nearest neighbors of
a newly attached particle, but that it is necessary to recheck the complete
set of sites that are connected to infinity. We perform this by an
algorithm that marks this set recursively.  

Special care
has to be taken for upward growth processes, because they may cause $\Delta m$
to exceed $O$. In case this happens, the bottom row of the configuration is
truncated. Finally, symmetry has to be taken into account. Rotations around
the axis of the cylinder and reflections about any vertical axis
do not change the
growth probabilities or the steady-state weights, so the set of all of 
the symmetric configuration are represented by a single canonical choice.
More specifically, a configuration is represented by a binary word that
consists of $N\times O$ digits that correspond to the sites; 
The empty sites that are on the exterior
are given a value of $1$, and the rest of the sites are assigned with zeros.
We choose the canonical form as the word that has the maximal numerical value.

After the complete list of configurations is processed, the calculation of
the evolution matrix is completed, and it is closed, i.e., $\sum_iE_{i,j}=1$
for every $j$. Then the steady-state vector ${\bf P}^*$ is calculated
iteratively
by applying the evolution matrix many times on some initial state vector.
This method is much faster than any of the standard techniques for solving
a set of linear equations, especially when the number of variables is
very large.
The next step is to calculate the average upward growth probability, according
to Eq. (\ref{pup=}), and the average density, according to Eq. 
(\ref{rho(N)=1/Npup}). Our computer resources enabled us to conduct the 
enumeration only up to a finite order $O_{\rm max}$ that depends on $N$.
As explained above, for $N=2$ and $N=3$
there exists a finite number of configurations, and 
higher order approximations are irrelevant.
One may be surprised that we are able to reach $O=8$ for $N=6$, but do not
reach such a high order for $N=4$ and $N=5$, because for sure, there are
less configurations in the same order of approximation for lower $N$'s. The
reason is that very good convergence is achieved already for $O=N$, so we
had little to gain by going to much higher orders, and we stop at
$O=N+2$ for $N=4$ and $N=5$. The calculated densities 
$\rho_c(N,O)$ are presented in Table \ref{EnumerationtResultsTab},
together with the number of configurations $N_c$. 
The Table also presents the 
extrapolation and simulation results. 
\subsection{The extrapolation of the order of approximation to infinity,
$O\to\infty$}
\label{OtoInfinitySec}
Very good accuracy (about $10^{-4}$) is also obtained for $N=8$, 
even-though $O_{\rm max}=N-2=6$.
However, for $N\geq 9$ the results are
not very accurate, because the maximal available order is only $O=5$
for $N=9,10$ and $O=4$ for $N=11,12$. In spite
of that, we are able to arrive at a more precise estimation for
$N=9,10$ by extrapolating $O\to\infty$. The extrapolation does not improve
the accuracy of the cases $N=11,12$ to a satisfactory level.
Our aim is to deduce the value of 
$\rho(N)=\lim_{O\to\infty}\rho_c(N,O)$ from the limited range of available 
values for $O$. We start by noting that our data practically
reached asymptotia for 
$N=4,5,6$. We detect that the differences, $\rho_c(O,N)-\rho_c(O+1,N)$,
decay exponentially and thus conclude that the function
$f=\ln\left[\rho_c(N,O)/\rho(N)-1\right]$, is very close to being linear. 
Substituting the parameterization $f=\beta-\alpha O/N$ we are able to extract
the three unknowns, $\alpha$, $\beta$, and $\rho(N)$ using at least three
data points.
For $N=6$ and $O=4,5$, and $6$, we find that $\beta=0.03$ and $\alpha=12.31$.
The value of $\rho(6)$ turns out to be very close to the highest available
approximation $\rho_c(6,8)$. 

Scaling theory would imply that, for large $N$ and $O$, $f$ should become
a universal function, which depends only on the scaled ratio $x=O/N$
(without an additional dependence on $N$). Following this expectation, we
thus conjecture the general relation
\begin{equation}
\rho_c(N,O)=\rho(N)\left[1+e^{f(O/N)}\right],
\end{equation}
with $f(x)\simeq -12.3x$, for $N,O\gg 1$.

\begin{figure}
\epsfbox{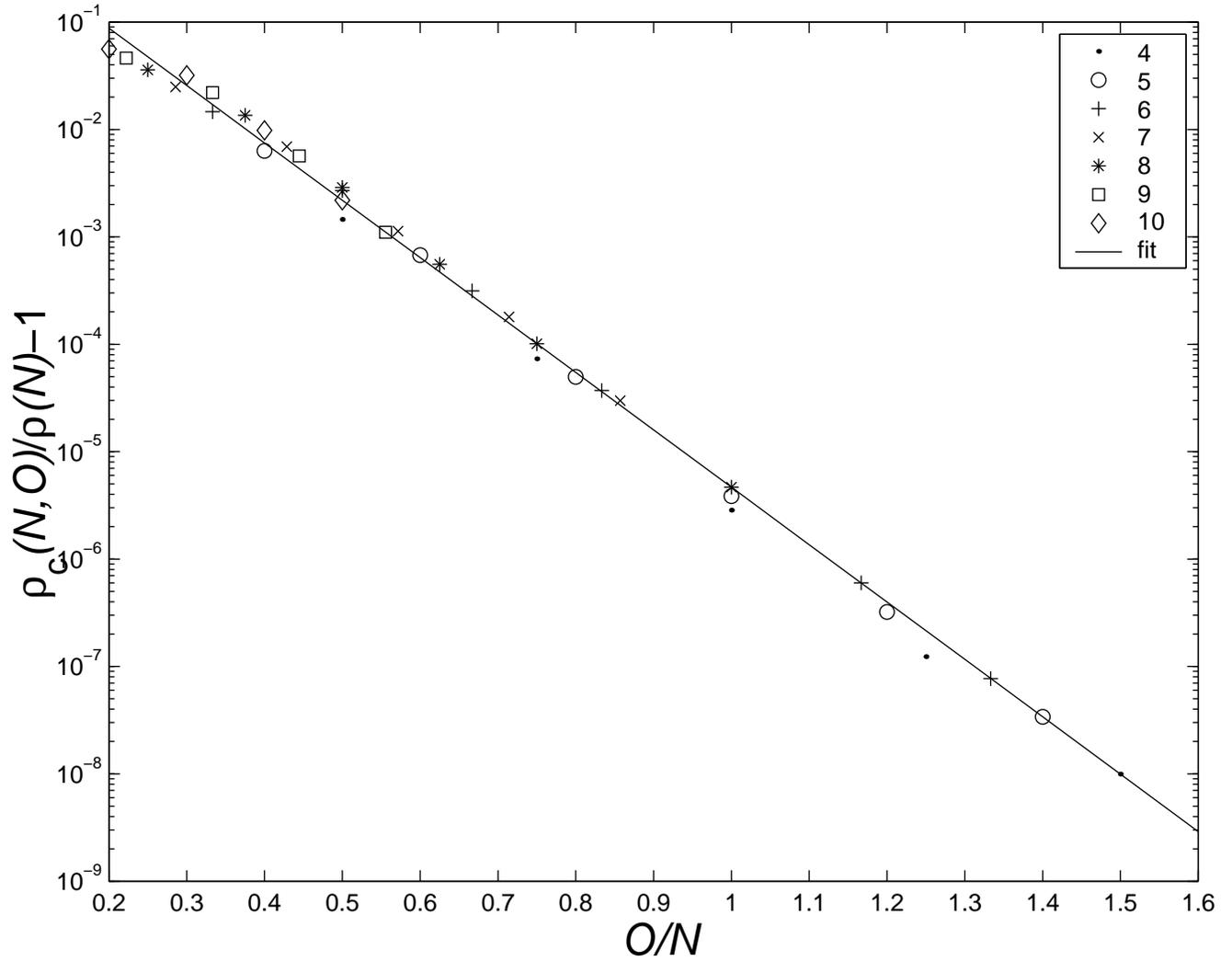}
\vspace{5mm}
\caption{Data collapse of $e^f=\left[\rho_c(N,O)/\rho(N)-1\right]$ vs. $O/N$ 
for all the data points with $4 \leq N \leq 10$ and $O>2$, 
on a semi-log scale. The continuous line shows the linear 
approximation $f \simeq -12.3O/N$.}
\label{DataCollapseFig}
\end{figure}

To test this conjecture, we estimated $\rho(N)$, for $N\geq 4$, via
\begin{equation}
\rho(N)\simeq\frac{\rho_c(N,O_{\rm max})}{1+e^{f(O_{\rm max}/N)}}.
\label{XX}
\end{equation}
We have then used this estimate to calculate $\rho_c(N,O)/\rho(N)$ for
$O<O_{\rm max}$. The resulting values are shown in Fig. 
\ref{DataCollapseFig}, together with the line $f(x)=-12.3x$. 
Clearly, all the values for $O/N \gtrsim 0.4$ are
consistent with our conjectured form for $f(x)$.

The values of $\rho(N)$, as deduced using Eq. (\ref{XX}), are listed in
Table \ref{EnumerationtResultsTab}. Clearly, they all agree with the values
from the simulations, except for small deviations that appear for $N=9$ and
$10$. In the cases $N=11,12$ the deviations are relatively large, because
$O_{\rm max}$ is too small, and hence the extrapoltion results are not
specified.

\subsection{An enumeration based estimate of the fractal dimension $D$}
\label{EnumeartionD}
In the previous section, we
obtain very accurate estimates of the asymptotic ($O\to\infty$)
average steady-state densities $\rho(N)$. In this section, we 
extrapolate the latter densities in the limit $N\to \infty$, in
order to find the fractal dimension $D$.
Consider a $N^d$ segment in the steady state regime of growth. Assuming
that the structure is a self similar fractal, which has no
characteristic length scale other than $N$, we expect that the average mass
of the segment would be proportional to $N^D$, and that the density
would be proportional to $N^{D-d}$. 
In principle however, 
one expects some corrections to scaling  
as in Eq. (\ref{rho(N)=AN^(D-d)}). Taking only the first correction
term of that equation into account 
we get an approximation that depends on four parameters:
$D$, $A$, $B$ and $\theta$:
\begin{equation}
\rho_a(N)=AN^{D-d}\left(1+\frac{B}{N^{\theta}}\right),
\label{rho(N)approx1}
\end{equation}
where the subscript $a$ denotes that this is an approximation.
Using the four data points with $7\leq N\leq10$, a fit to Eq. 
(\ref{rho(N)approx1}) yields $D=1.64$, $\log(A)=-0.63$, $B=1.31$ and
$\theta=1.48$.
The calculation of the parameters can also be based 
on more than the minimal four points, using a least mean
square error method. We choose to minimize the logarithmic (or relative) 
errors $\Delta\rho/\rho$ rather than the errors in the densities $\Delta\rho$,
because we find them to be more uniformly distributed.  
The results of the fit using the six data points with $5\leq N\leq10$ yields
$D=1.74\pm0.06$, $\log(A)=-1.0\pm0.3$, $B=1.5\pm0.6$, and 
$\theta=0.80\pm0.13$. The error estimates are evaluated using a 
confidence level of $0.95$.
Since the fit yields a value for $\theta$ that is
close to $1$, we also try a three parameter fit, fixing $\theta=1$. Using the
three rightmost data points, for $N=8,9$ and $10$, gives $D=1.68$,
$\log(A)=-0.799$, and $B=1.16$. Using more points with $5\leq N \leq 10$,
we get
\begin{eqnarray}
&&D=1.68\pm0.01 \nonumber \\
&&\log(A)=-0.784\pm0.016 \nonumber \\
&&B=1.12\pm0.05.
\end{eqnarray}

Finally, an alternative four parameter form, including only ``analytic''
corrections, is 
\begin{equation}
\rho_a(N)=AN^{D-d}\left(1+\frac{B}{N}+\frac{C}{N^2}\right).
\label{rho(N)approx2}
\end{equation}
This time the results for $7\leq N\leq 10$ are $D=1.65$, $\log(A)=-0.68$,
$B=0.55$ and $C=1.10$, and the least mean square calculation for $5\leq
N \leq 10$ yields
$D=1.70\pm0.02$, $\log(A)=-0.87\pm0.08$, $B=1.5\pm0.4$, and 
$C=-0.5\pm0.4$.

We thus conclude that the fractal dimension of cylindrical DLA is 
$D\simeq 1.68\pm0.01$, close to the results of
earlier numerical work \cite{Erzan95}.
\section{Simulation}
\label{SimulationSec}
As mentioned, our analytical enumeration results are confirmed by simulations.
In this section we describe how our simulations were conducted, with special
attention to the boundary Green's function $g_N(n)$, which
is given a new probabilistic meaning. We also discuss
the accuracy of the results, and finally, we try to fit the results
to some approximations as in the end of the previous section 
and obtain some more estimates of the fractal dimension. 

Our simulation is performed on a lattice, which is represented by a
2D array variable. Each of the variables in the array can assume one of two 
possible values, $1$ or $0$, that determine whether the 
relevant site is occupied by an aggregate particle or not, respectively.
The size of the array is $(14N)\times N$, i.e.,
its width is $N$ and it is composed of $14$ blocks of $N\times N$ sites
stacked one on top of the other. The number $14$ is quite arbitrary and
could be chosen differently. In principle, the lattice should
be tall enough to allow the aggregate to arrive at a steady state, and also
to allow a margin at the top, because the average density of the aggregate
is lower near the growing front. 
Each time a new cluster is initialized,
the lattice array is cleared so that all of its variables are set to
$0$, except for the bottom row, which is set to $1$. This means
that the initial shape of the aggregate is a horizontal line at the
bottom of the lattice.
A random walker is characterized by the coordinates
$(m,n)$ of its position. In each simulation step a direction is chosen 
randomly and the particle is advanced in that direction. If the particle
happens to go into a site that is nearest neighbor to the aggregate then
it sticks, i.e., the value of the relevant lattice variable is updated from
$0$ to $1$.  Then the next random walker is released, and so on.
\subsection{The role of the Green's function}
\begin{figure}
\epsfbox{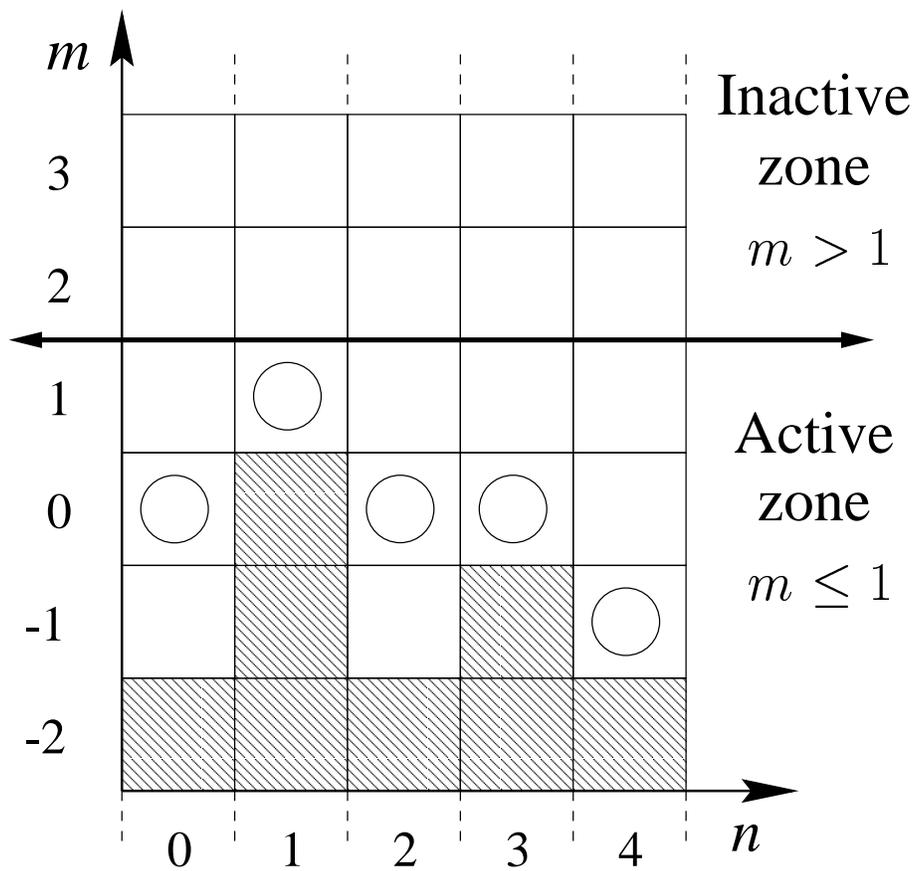}
\vspace{5mm}
\caption{The bold line separates between the upper inactive zone with $m>1$
and the lower zone with $m\leq 1$. A random walker cannot stick 
in the inactive zone.}
\label{BoldLineFig}
\end{figure}

In principle, each new random walker should be released far above the 
aggregate, near the upper distant boundary. In practice, nothing can happen to 
the random walker (it cannot stick) 
until it crosses the bold line in Fig. \ref{BoldLineFig} 
This line is drawn between
the highest row where a random walker can stick ($m=1$) and the row above it
($m=2$), and thus
it differentiates between the active zone below the line, 
with $m \leq 1$, and the inactive 
zone above it, with $m>1$.
The projection of the path of the random walker on the vertical axis (its $m$
coordinate) is also a random walk, only in one dimension (1D).
Usually in 1D there is a probability of $1/2$ to
go up and the same probability to go down, but in our case, 
there is a probability of $1/4$ to go in either direction, and a probability
of $1/2$ to stay at the same row. Nevertheless, 
this motion is still equivalent to 
a random walk, however the effective time step is longer. A quality of
1D random walks is that there is a probability of $1$ to arrive at any site
(no matter how far) within a finite time. Therefore, 
there is a probability $1$ that eventually the random walker would cross
the line from the inactive zone into the active zone.
The random walker is equally probable to cross this line at any of the $N$ 
sites, so instead of waiting for a long time, 
it is  more efficient to start the simulation by
inserting the random walker in a random site just below the line in the
active zone \cite{Kaufman95}.

But what happens if the path of the random walker happens 
to cross the line into the inactive zone? Once more we apply
the same reasoning and claim that ultimately the random walker would re-cross
the line downwards at some point with probability $1$. Unlike the initial
insertion, this time the distribution of the reentry point is not uniform.
It is quite easy to see for example, that there is a greater chance for the 
particle to reenter at the exact same site from which it exited than 
for it to reenter at a site that is far away. Let us denote 
by $\Psi(m,n;n')$ the
probability that if the particle is at some initial site $(m,n)$ in the 
inactive zone ($m>1$), it will cross the
line for the first time at $(m'=1,n')$. In the next
time step the random walker moves to one of its nearest neighbors with
equal probability. Therefore $\Psi(m,n;n')$ must be equal to the
average of $\Psi$ on all the nearest neighbors. This implies that $\Psi$
satisfies the Laplace equation (in the coordinates $m$ and $n$),
\begin{equation}
\nabla^2\Psi(m,n;n')=0.
\end{equation} 
The boundary conditions for $\Psi$ at the lower boundary are
\begin{equation}
\Psi(m=1,n;n')=\left\{
\begin{array}{ccc}
1&,&n=n' \\
0&,&\mbox{otherwise}
\end{array}
\right..
\end{equation}
This is true because if the
random walker is already in the row $m'=1$ then it already passed the line
between $m=1$ and $m=2$ and so it stops before it starts. 
The boundary conditions at the
top are $\Psi={\rm const.}$, or equivalently 
\begin{equation}
\lim_{m\to\infty}\frac{\partial\Psi}{\partial m}=0.
\end{equation}
These are the exact same conditions satisfied by the Green's function
\cite{Kol98},
and so the theorem about the uniqueness of the solution of the Laplace equation
with boundary condition assures that $\Psi$ is equal to the Green function,
and especially at the first row above the line $m=1$,
\begin{equation}
\Psi(1,n;n')=g_N(n-n').
\end{equation}

This means that each time the random walker attempts to cross the line to the
inactive zone, it can be returned to the active zone immediately. 
The distance of the reentry point
from the exit point should be chosen randomly from the distribution defined
by the Green's function $g_N(n)$. This policy saves a lot of simulation time
in comparison with the alternative option of letting the random walker 
wander freely until it finally sticks, or until it passes some arbitrary 
critical distance from the aggregate. 
We note in passing that the discussion in this section proves
Eq. (\ref{sumg(n)=}) in an 
alternative, probabilistic approach, simply due to the fact that $g_N(n)$ is 
a probability function. 

\subsection{Analyzing the statistics}
A single cluster is completed as
soon as the first particle sticks in the top row of the lattice.
Then, the number of particles in each row is counted and
stored in a (1D) array variable that
represents the average {\it density profile} as a function of height.
Then the lattice array is
cleared and a new aggregate is started. In contrast, the density profile
array is not cleared, and it accumulates
data for each new cluster so that after many iterations it converges
to the average density, when normalized by the number of iterations $N_i$.

\begin{figure}
\epsfbox{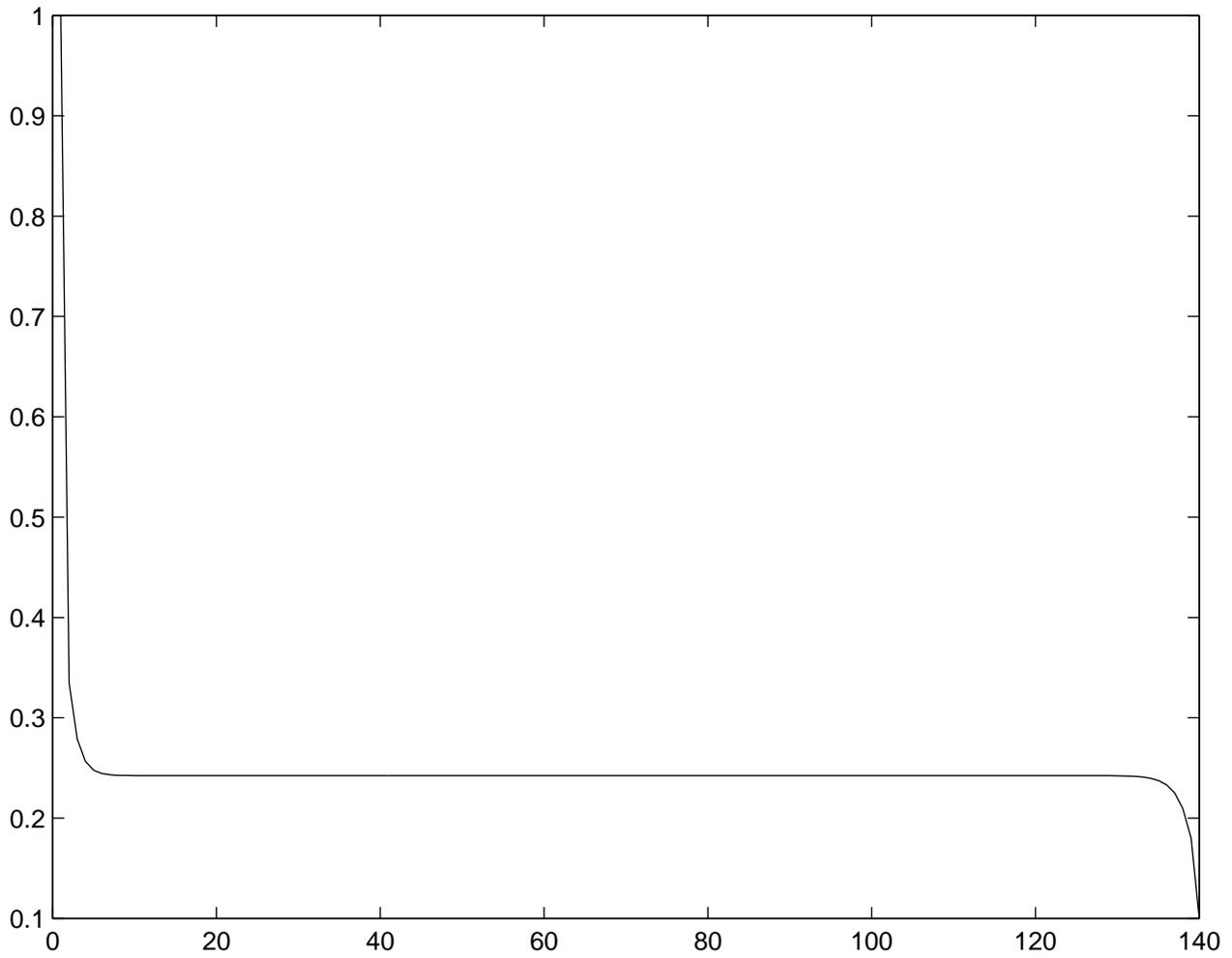}
\vspace{5mm}
\caption{The density profile as a function of height for a lattice with 
$N=10$ averaged over some $N_i\approx 2\times 10^7$ iterations. 
The height of the lattice is $140$ sites.}
\label{DensityProfileFig}
\end{figure}

An example of a density profile is shown in Fig. \ref{DensityProfileFig}, 
where $N=10$ and $N_i\approx 2 \times 10^7$. Three distinct regions are
visible in the graph; On the left part there is a fast decay from an initial
density of $1$ to a plateau. These graphs always start from a density of $1$,
because the initial conditions for growth are that the bottom row of the 
lattice is completely occupied. The decay to the plateau shows the convergence
to the steady state stage of the growth. It seems that the steady state
settles roughly at a height that is equal to the width, i.e., about $10$.
The middle section of the graph seems to be a flat plateau of constant density.
In fact, there are small statistical fluctuations due to the randomness of the
simulations, which are invisible because they are on the order of $10^{-5}$.
Finally, near the right end of the 
graph there is a decay to $0$. This is because the density near the 
growing front is smaller than in the frozen part, 
because more particles are still expected to stick there and finally raise
the density to the steady state value. Naturally, the density is $0$ above
the highest particle in the aggregate, which is always at the top row of the
lattice, because the simulation always stops at that point. It seems that
the width of the interface layer is also close to $N$. 

Only the middle section of the
aggregate is taken into account in measuring the average steady state density
$\rho(N)$. As mentioned, our
impression is that a margin of $N$ sites from each side is enough, but we
work with margins of $2N$. We thus evaluate $\rho(N)$ as the average of
the density profile over the plateau area. A possible
way of estimating the accuracy of this average is by taking the standard
deviation and normalizing by the square root of the number of rows that 
participate in the averaging. This is however, somewhat optimistic and
would produce very low error estimates, because this calculation assumes
statistical independence between adjacent rows, where in fact, there are
significant correlations. The right factor to normalize by is therefore the
square root of the effective number of independent rows. Since the aggregate
is fractal, the only available length-scale is $N$, and therefore the
correlation length $\xi$ should be proportional to $N$.
We therefore conservatively
guess that two rows that are $N$ rows apart are independent, and hence
estimate the accuracy by the standard deviation divided by $\sqrt{10}$, 
because there are $10$ blocks of $N\times N$ is the steady state region.
This error estimate is based on the expected dependence on the number of 
blocks, but there could be some numerical factor missing.

An alternative way of measuring $\rho(N)$ is by measuring $\left<p_{\rm up}
\right>^*$ directly and using Eq. (\ref{rho(N)=1/Npup}).
After the aggregate reaches a height of $2N$ we
assume that it is in the steady state and we start gathering statistics.
In particular, we count the number of upward growth events, when
the random walker sticks above all the particles in the aggregate.
Our results
show very good correspondence between the two different ways;
The typical relative difference is on the order of $10^{-6}$.

The simulations were carried out for the following values of $N$: 
$2,3,\dots,12$, $16$, $24$, $32$, $48$, $64$, $96$ and $128$. We
did not go beyond that because our computer resources
did not suffice to iterate a large enough number of clusters to obtain a
relative accuracy of about $10^{-4}$ or better, as obtained for the
other cases. 

\begin{figure}
\epsfbox{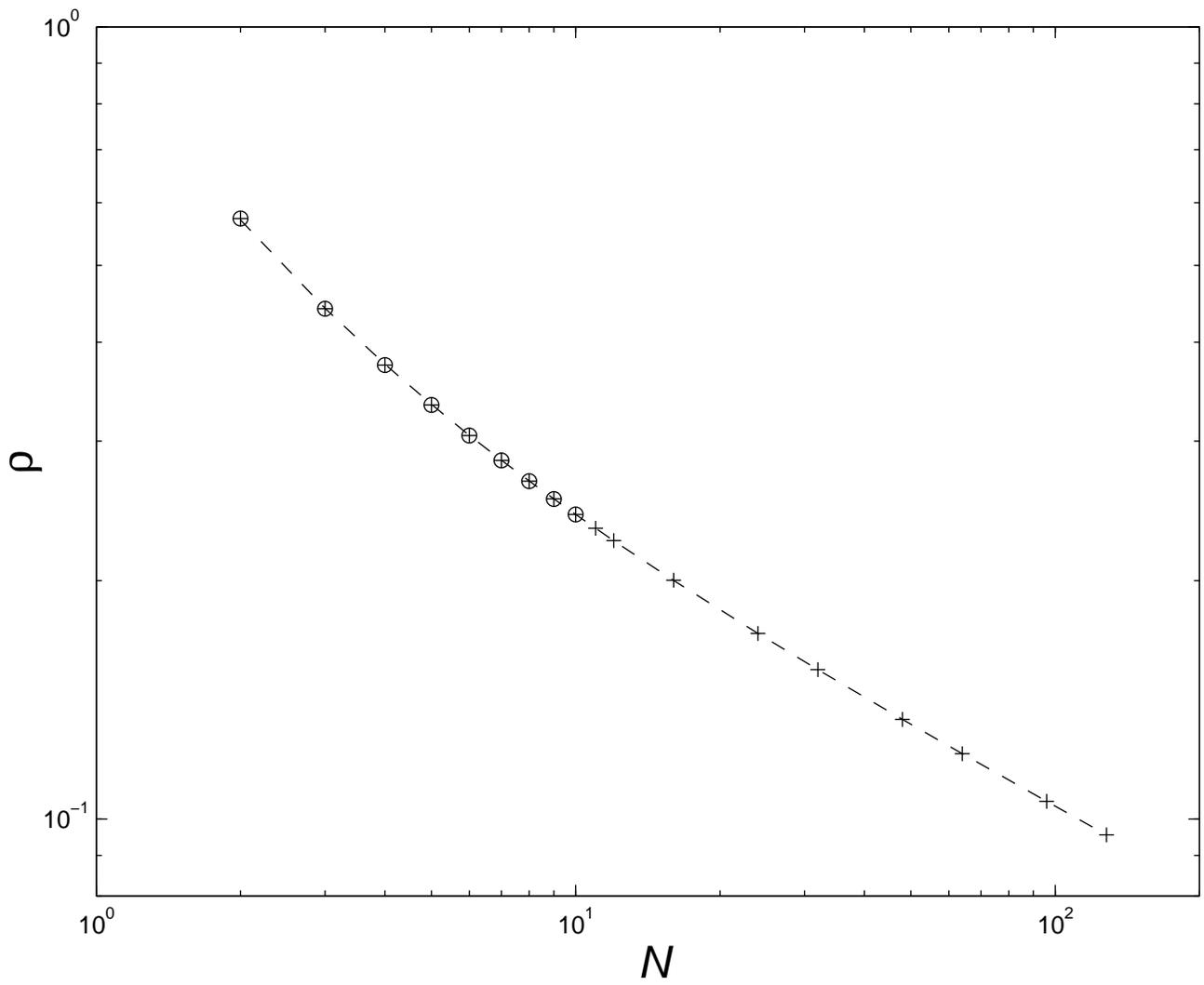}
\vspace{5mm}
\caption{A plot of $\rho(N)$ vs. $N$ on log-log scales. 
The plus signs denote the simulation
results, the dashed line denotes $\rho_a(N)$ - the best fit to the three 
parameter approximation of Eq. (\protect\ref{rho(N)approx1}), 
with $\theta=1$. The circles denote the enumeration results.}
\label{RhovsNFig}
\end{figure}

\begin{figure}
\epsfbox{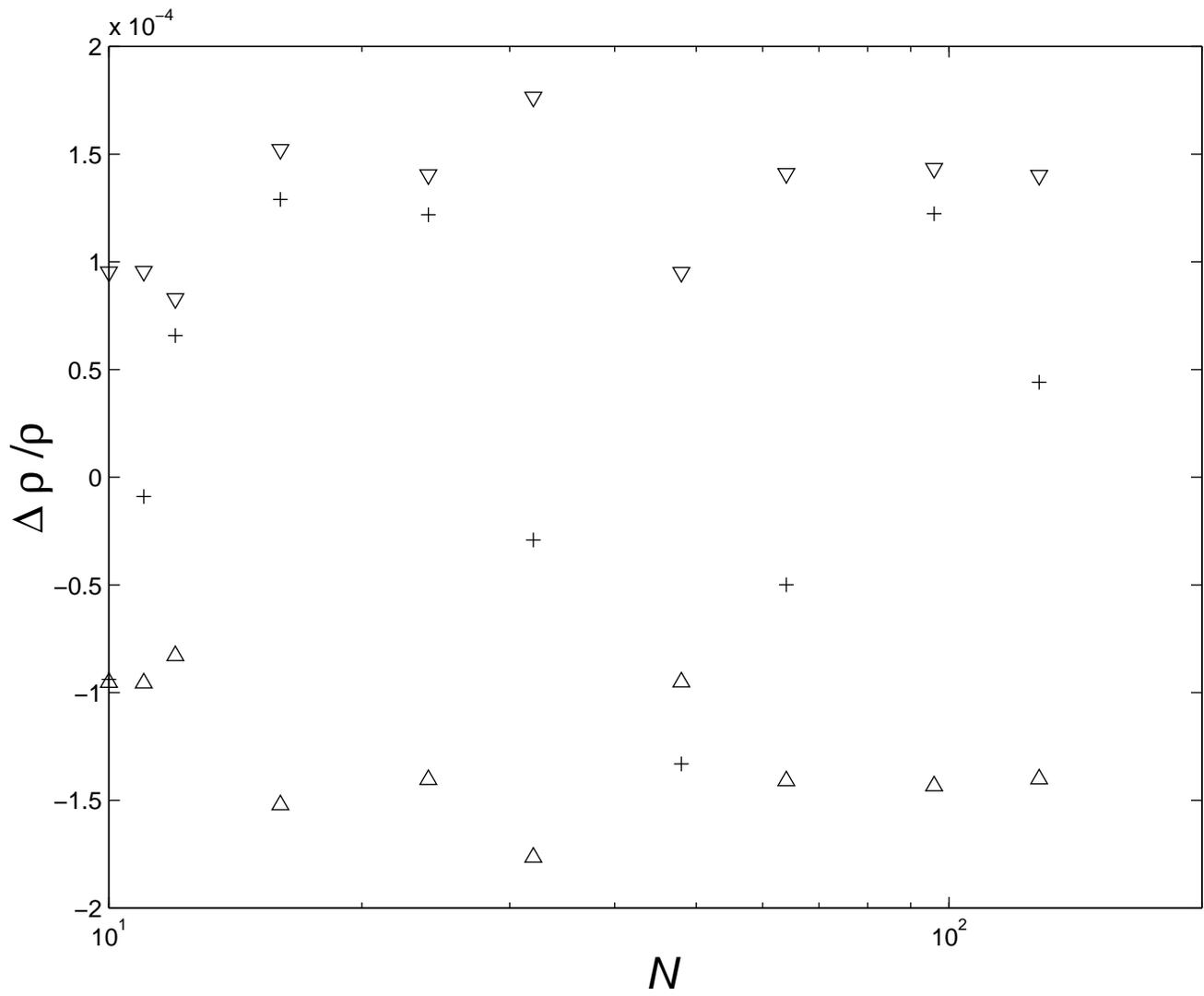}
\vspace{5mm}
\caption{The relative residuals $v=(\rho-\rho_a)/\rho$ 
(the plus signs) vs. $N$ on semi-log scales. The upper and lower triangles
show the estimated confidence intervals (errors) of the simulation data
,$\pm\sigma$.}
\label{ResidualsFig}
\end{figure}

We now proceed to fit the results for the $10$ available data points with 
$128 \geq N \geq 10$ in a similar way to Sec. \ref{EnumeartionD}.
The difference is that now we use the error estimates $\sigma_i$ to
give weights to the different data points, because not all the accuracies are
the same. This way the fit will allow greater residuals for data points with
larger error estimates. 
Our first attempt is to fit the four parameter approximation of Eq. 
(\ref{rho(N)approx1}). The results are $D=1.673\pm0.002$, 
$\log(A)=-0.770\pm0.0013$, $B=1.03\pm0.06$, and $\theta=0.96\pm 0.06$.
The maximal relative residual is $1.2\times 10^{-4}$. 
Once more we set $\theta=1$ and perform the fit for the remaining three
parameters. The results are 
\begin{eqnarray}
&&D=1.671 \pm0.001, \nonumber \\ 
&&\log(A)=-0.762\pm0.003, \nonumber \\
&&B=1.071\pm0.015.
\end{eqnarray}
The resulting error estimates seem a bit too optimistic, perhaps
also becuase of the presence of some systematic errors that are
not taken into account.
The simulation results are shown in Fig. \ref{RhovsNFig} 
on log-log scales as
plus signs, along with the latter three-parameter fit,
shown as a dashed line. The figure also shows the enumeration results as
circles. 
Since the differences are hardly noticeable, we display the relative
(logarithmic)
residuals $v_i\equiv\left(\rho(N_i)-\rho_a(N_i)\right)/\rho(N_i)$ 
separately in Fig. \ref{ResidualsFig} on semi-log scales, in
comparison with the relative error estimates $\pm \sigma_i$. 
The maximal relative residual is $1.3\times 10^{-4}$. This
is consistent with the order of magnitude of the estimated a priori errors.

A factor that indicates the compatibility between 
the a priori error estimates $\sigma_i$ and the a posteriori residuals $v_i$
is,
\begin{equation}
\chi^2\equiv\frac{1}{N_d}\sum_i\left(\frac{v_i}{\sigma_i}\right)^2,
\end{equation}
where $N_d$, the number of degrees of freedom, is equal
to the number of data-points minus the number of unknown parameters.
The value of $\chi^2$ should be close to $1$.
In the latter fit we get $\chi^2=0.9$, whereas $\chi^2=0.3$ in the former.
The results of the fit imply that the three parameter approximation is sound.

For the sake of comparison we also try to fit to the other test approximations 
that were introduced in the previous section. The best fit to the 
four parameter approximation
in Eq. (\ref{rho(N)approx2}) is 
$D=1.6721\pm0.0012$,
$\log(A)=-0.766\pm0.006$,
$B=1.12\pm0.07$ and $C=-0.28\pm0.4$. In this approximation,
the residuals are not lowered drastically; the maximal relative
residual is $1.1\times10^{-4}$ and $\chi^2=0.2$.
The error estimate of fourth parameter $C$ is much greater than the error 
estimates of the other parameters. 
The contribution of the term with the
$C$ parameter is on the same order of magnitude as the residuals, at least
for the data point with large $N$'s. This implies that there may be significant
contributions from the noise (the errors) in these data points to the
parameter, and therefore its inclusion is redundant. 

\section{Summary}
\label{SummarySec}
In this paper, we continue our endeavour to solve cylindrical DLA 
analytically, i.e, to calculate
the steady state average density $\rho$, as a function of the cylinder width 
$N$, and to find the fractal dimension $D$. Unlike our previous work, which
deals with bond-sticking conditions \cite{Kol2000a}, this work solves
for site-sticking conditions. 
The immediate problem in following our Markovian method is that,
except for $N=2,3$, there is 
usually an infinite number of configurations.
The case $N=4$ has an infinite
number of configurations, but is still relatively simple.
The large variety of possible complex interface shapes for $N\geq 5$ 
prevents the
inclusion of all the configurations and compels the use of an approximation
scheme, in which only a finite number
of rows $O$ of the growing front near the tip are included.
This approximation works
because of the exponential decay of the Laplace potential $\Phi$ inside
deep fjords.
The approximation
leaves a finite number of configurations to work with, and thus
the computational procedure can be completed.

We find that this
is a well controlled approximation, in the sense that any
desired numerical accuracy can be achieved provided that a high enough order of
approximation $O$ is used. The results are summarized in Table 
\ref{EnumerationtResultsTab}, that shows the computed density $\rho_c$ for
various values of $N$ and $O$ along with the number of 
relevant configuration $N_c$. An evident fact is that $N_c$ grows very
rapidly as a function of $O$ and $N$, making it impractical to perform
the calculation for wide cylinders. We note that in order to obtain
the same relative accuracy it is necessary to use $O\propto N$, e.g., in
order to obtain a relative accuracy better than $10^{-4}$ one should use
at least $O=N-1$. This is the case for $N\leq 7$, where the results are very
accurate, but not so for $N\geq 8$, where our available computer
resources allowed only lower order computations. 
As discussed in Sec. \ref{OtoInfinitySec}, we are able to improve
the estimates in these cases by extrapolating $O\to \infty$,
taking advantage of the universal exponential decay of 
$\rho_c(N,O)/\rho(N)$ with the scaled variable $O/N$.
Table \ref{EnumerationtResultsTab} also compares the enumeration estimates 
with direct measurements from simulations, and finds them to agree within
the simulation errors.
Once accurate estimates are obtained for $\rho(N)$ for $N \leq 10$, they
are fitted to a power-law approximation with a correction to scaling term 
according to Eq. (\ref{rho(N)approx1}).
The fit (with $\theta=1$) gives an estimate of the fractal 
dimension $D=1.68\pm0.01$.

Besides the range $2\leq N \leq10$, simulations are also 
performed on cylinders with
larger $N$'s in the range $10 \leq N \leq 128$. The relative errors of
the measurements of $\rho(N)$ are estimated around $10^{-4}$. 
The simulation data are also fitted to the same approximations. Once again, the
three parameter approximation proves most appropriate and the resulting fractal
dimension this time is $D=1.671$. The fact that the enumeration
and simulation based estimates of the fractal dimension are very close
is a good indication of their accuracy.

The last statement should be taken with some caution in light of evidence
that raises doubts concerning self-similarity in radial DLA 
\cite{Mandelbrot92,Mandelbrot95a,Mandelbrot95b}, or suggesting
some very slow crossovers \cite{Somfai99,Mandelbrot2000}. 
Indeed, radial
DLA is somewhat different than cylindrical DLA, as manifested by
the difference between their fractal dimensions: $D=1.71$ for 
radial DLA \cite{Tolman89,Ossadnik91} and $D=1.66$ for cylindrical DLA
(this difference is still not fully understood).

We also tried performing the exact
calculations for $N=11$ and $12$, but managed to
go only up to $O_{\rm max}=4$. This was insufficient for extrapolation with
an accuracy that is comparable to the rest of the data points. With the aid
of stronger computers we think that it would be possible and beneficial 
to compute a few more data points $\rho(N)$, which would help obtaining more
accurate estimates of the fractal dimension. Also, the techniques discussed
here could be used to find the fractal dimension of cylindrical DLA in 3D.
However, since a much larger number of configurations can be expected, this
task would also probably require the aid of a very strong computer.

There are a few differences between site-DLA and bond-DLA: The boundary
conditions for the Laplace equation are a little bit different; 
In bond-DLA the potential is set to
zero on the aggregate itself, whereas in site-DLA the potential is set to
zero on sites that are nearest neighbors of the aggregate. Also, the
growth probabilities are computed somewhat differently; In bond-DLA
contributions are summed over bonds that go out of a site
where sticking may occur, whereas they
are summed over bond that go into it in site-DLA. The normalization factor,
however, is equal to the width $N$ is both cases. In the case of site sticking
conditions there is an effective thickening of branches and thus a narrowing
of fjords. 
Thus, there is a notable decrease in the probability of a random
walker to penetrate deep into fjords. 
This also causes the number of configurations for a particular choice of $N$
and $O$ to be considerably less for site-DLA in comparison with bond-DLA.
Therefore, accurate enumeration 
results can be obtained for larger $N$'s and $O$'s in site-DLA. 
The extrapolation $O\to\infty$ performed in this paper was not done
in Ref. \cite{Kol2000a}, which deals with bond-DLA, because the technique was
not developed at that time. When we apply the method to the bond-DLA case,
we manage to improve the relative accuracy of the highest available 
approximations, $\rho_c\left(N,O_{\rm max}(N)\right)$
for $N=6,7$ by an order of magnitude: from about
$1.2\times10^{-3}$ to $2\times10^{-4}$ for $N=6$, and from $5\times10^{-2}$
to $1.6\times10^{-3}$. This extrapolation is based on the data points
for $N=5$.
The relative accuracy of 
$\rho_c\left(N,O_{\rm max}(N)\right)$ for $N\leq 5$ is better than $10^{-4}$
and hence, the extrapolation is not necessary.
The estimate of the fractal dimension for site-DLA is,
$D=1.68$, to be compared with the bond-DLA enumeration result $D=1.64$ 
\cite{Kol2000a}. In contrast, the difference in the simulation results for 
the two cases is smaller: $D=1.67$ for site-DLA and $D=1.66$ for bond-DLA.
Given the uncertainties, our results are consitent with 
universality with respect to
the sticking conditions \cite{Cafiero93,Erzan95}.
\acknowledgments
We thank Peter Jones, Brooks Harris, Yonathan Shapir, Torstein J{\o}ssang,
Barbara Drossel, and Leonid Levitov for helpful
discussions. We also thank Yiftah Navot for helping with the
computer program, by suggesting more efficient data structures and algorithms.
This work was supported by a grant from the German-Israeli Foundation (GIF).

\end{multicols}
\widetext

\newpage
\begin{table}
\caption{The approximated densities $\rho_c$ and the number of configurations 
$N_c$ for various orders $O$ and cylinder widths $N$. The approximated 
densities from enumeration are compared to simulation results. In addition,
the extrapolated density $\rho(N,O\to\infty)$ is also presented.}
\begin{tabular}{c|cccccccccc}
$N/O$&simulation&$O\to\infty$&1&2&3&4&5&6&7&8\\
\tableline
2&0.5732&& 0.5732 \\
&&& 2 \\

3&0.4408&& 0.4408 \\
&&& 3 \\

4&0.3744&0.3744&0.3743&0.3750&0.3744&0.3744&0.3744&0.3744\\
&&&5&6&7&8&9&10\\

5&0.3334&0.3334&0.3323&0.3355&0.3336&0.3334&0.3334&0.3334&0.3334\\
&&&7&10&14&24&52&134&378\\

6&0.3049&0.3049&0.3025&0.3094&0.3057&0.3050&0.3049&0.3049&0.3049&0.3049\\
&&&12&21&35&94&395&1970&10344&55161\\

7&0.2837&0.2837&0.2798&0.2908&0.2857&0.2840&0.2837&0.2837&0.2837\\
&&&17&38&76&280&1831&13575&98479\\

8&0.2671&0.2671&0.2616&0.2767&0.2707&0.2679&0.2672&0.2671\\
&&&29&81&190&846&7605&83043\\

9&0.2536&0.2537&0.2467&0.2655&0.2593&0.2551&0.2540\\
&&&45&161&451&2421&29220\\

10&0.2424&0.2426&0.2341&0.2562&0.2503&0.2450&0.2431\\
&&&77&349&1152&7213&111067 \\

11&0.2329&&0.2233&0.2483&0.2431&0.2368\\
&&&125&733&2885&21688\\

12&0.2247&&0.2139&0.2415&0.2371&0.2300\\
&&&223&1627&7504&67450

\end{tabular}
\label{EnumerationtResultsTab}
\end{table}
\end{document}